\documentclass[aps,prd,reprint,amsmath,amssymb,superscriptaddress,floatfix]{revtex4}
\usepackage{graphicx}
\usepackage{amsfonts}
\usepackage{amssymb}
\usepackage{rotating}
\usepackage{booktabs}
\usepackage{xcolor}
\usepackage{soul}
\usepackage{color}
\usepackage{slashed}
\usepackage{multirow}
\usepackage{makecell}
\usepackage{epsf}
\usepackage{ulem}
\usepackage{cancel}
\usepackage{color,bm}
\usepackage[colorinlistoftodos]{todonotes}

\usepackage[colorlinks=true,citecolor=cyan,urlcolor=blue,bookmarks=true,bookmarks=true,bookmarksopen=true,bookmarksnumbered=true,bookmarksopenlevel=3]{hyperref}

\definecolor{airforceblue}{rgb}{0.36, 0.54, 0.66}
\definecolor{steelblue}{rgb}{0.27, 0.51, 0.71}
\definecolor{amber}{rgb}{1.0, 0.49, 0.0}

\begin{document}

\title{Transverse single spin asymmetry $A_{UT}^{\sin(\phi_h-\phi_S)}$ for single hadron production in semi-inclusive
 DIS}
\author{Xuan Luo}
\author{\textsc{Hao Sun}\footnote{Corresponding author: haosun@mail.ustc.edu.cn \hspace{0.2cm} haosun@dlut.edu.cn}}
\affiliation{ Institute of Theoretical Physics, School of Physics, Dalian University of Technology, \\ No.2 Linggong Road, Dalian, Liaoning, 116024, People’s Republic of China }
\date{\today}

\begin{abstract}
In this paper, we study the single spin asymmetry $A_{UT}^{\sin(\phi_h-\phi_S)}$ of a single hadron production in semi-inclusive deep inelastic scattering within the framework of transverse momentum dependent factorization up to the next-to-leading logarithmic order of QCD.
The asymmetry is a contribution of the convolution of the Sivers function and the unpolarized fragmentation function.
Specifically, the Sivers function in the coordinate space and perturbative region can represent the convolution of the $C$ coefficients and the corresponding collinear correlation functions, among which the Qiu-Sterman function is the most relevant one. We perform a detailed phenomenological analysis of the Sivers asymmetry at the kinematics of the HERMES and the COMPASS measurements. It is found that the obtained $x_B$-, $z_h$-, and $P_{h\perp}$-dependent asymmetries are basically consistent with the HERMES and the COMPASS measurements.
\vspace{0.5cm}
\end{abstract}
\maketitle
\setcounter{footnote}{0}

\section{introduction}

Since they were first observed, transverse single spin asymmetries (SSA) are a topic in spin physics of significant theoretical and experimental interest \cite{Adams:1991rw,Adams:1991cs,Arsene:2008aa}.
SSA appear in a scattering process when one of the colliding proton or the target is transversely polarized with respect to the scattering plane. They can provide information on the three-dimensional structure of the nucleons. From the theoretical point of view, to explain the SSA, one requires the nonperturbative correlators of a quark or gluon, and there are two methods for this purpose. The first one is the transverse momentum dependent (TMD) factorization approach \cite{Ji:2004xq,Ji:2004wu}, where the inclusive cross section is written as a convolution of transverse momentum dependent partonic distribution functions (TMD-PDFs), transverse momentum dependent fragmentation functions (TMD-FFs), and QCD partonic cross sections. This method is phenomenologically well studied in Refs. \cite{Ji:2004xq,Ji:2004wu,GarciaEchevarria:2011rb,Bacchetta:2006tn,Anselmino:2002pd,Boer:1999mm,Arnold:2008kf,Boer:1997mf,Anselmino:2007fs}. The second approach describes the SSA as a twist-3 effect in the collinear factorization and is suited for describing SSA in the large $p_T$ region. This formalism was originally proposed and further developed by Refs.\cite{Efremov:1984ip,Qiu:1998ia,Kanazawa:2000hz,Kouvaris:2006zy,Eguchi:2006mc,Kanazawa:2014dca}.

Among the single spin asymmetries, the Sivers asymmetry plays a vital important role. The Sivers function \cite{Sivers:1989cc}, contributing to the Sivers asymmetry, represents an azimuthal dependence on the number density of unpolarized quarks inside a transversely polarized proton. It has been found that the initial and the final state interactions (gauge links) contribute to the Sivers asymmetry significantly; therefore, the Sivers function is process dependent \cite{Boer:2003cm}. For example, the Sivers function probed in semi-inclusive deep inelastic scattering (SIDIS) are expected to be the same in magnitude but opposite in sign compared to the one probed in the Drell-Yan process. The Sivers asymmetry has been measured in SIDIS by HERMES \cite{Airapetian:2009ae}, JLAB \cite{Qian:2011py}, and COMPASS \cite{Alekseev:2008aa,Adolph:2012sp,Adolph:2014zba} experiments. To obtain reliable theoretical estimate of the Sivers asymmetry, the scale evolution effects should be included. Since most of the data from above experiments are at a low transverse momentum ($P_{h\perp}$) of the hadron, 
a natural choice for the analysis is the TMD factorization, 
which is valid in the region where the hadron $P_{h\perp}$ is much smaller than the hard scale $Q$.
The Sivers asymmetry in Drell-Yan process within TMD factorization has been studied in Ref.\cite{Wang:2018pmx}.

In Ref.\cite{Echevarria:2014xaa}, the authors study the Sivers asymmetry in SIDIS considering TMD evolution. 
Following factorization theorems, the so-called TMD evolution based on the previous works by Collins-Soper-Sterman (CSS) \cite{Collins:1981uk,Collins:1984kg}, has been well boosted in recent years. After working out the evolution equation, the evolution from one energy scale to another, is described by the Sudakov form factor \cite{Collins:1984kg,Collins:2011zzd,Collins:1999dz} which can be divided into a perturbatively calculable part $S_{\rm P}$ and a nonperturbative part $S_{\rm NP}$. Concretely, TMD evolution is performed in coordinate $b$ space which is related to momentum ($k_\perp$) space via a Fourier transformation. In $b$ space the cross sections can be expressed by simple products of $b$ dependent TMDs, in contrast to convolutions in momentum  space. Then the Sudakov form factor becomes nonperturbative at large separation distances $b$; while at small $b \ll 1/\Lambda_{\rm QCD}$, it is perturbative and therefore, can be worked out order by order in a strong coupling constant $\alpha_s$. The $b$ dependence of TMDs related to their collinear counterparts, such as collinear parton distribution functions (PDFs), fragmentation functions (FFs), or multiparton correlation functions, can be calculated in perturbation theory. Specifically, the Sivers function in the $b$ space and perturbative region can be represented as the convolution of the $C$ coefficients and the corresponding collinear correlation functions. Among different collinear correlation functions, the Qiu-Sterman function $T_{q,F}(x,x)$ appearing in the structure function $\widetilde{F}_{UT}^\alpha(Q,b)$ (introduced in Sec.\ref{secII}) at the leading order, is the most relevant one \cite{Kang:2011mr}. Other twist-3 correlation functions that appear in the next-to-leading order corrections are ignored in this paper. In order to get trustworthy results, in this paper we consider the perturbative Sudakov form factors and the $C$ coefficients up to the next-to-leading logarithmic (NLL) accuracy. We perform the TMD evolution to reach the fragmentation function and Qiu-Sterman function at an initial scale $\mu_b=c/b^*$ by the evolution package QCDNUM \cite{Botje:2010ay}. These are different from Ref.\cite{Echevarria:2014xaa} where the authors approximately adopt the Qiu-Sterman function at $Q_0=2.4$GeV and the C coefficients up to leading order. Considering all the details above, in this paper we estimate the Sivers asymmetry within the TMD factorization and provide some updated phenomenological applications. Typically, we also have compared the results with the HERMES and COMPASS measurements.

The paper is organized as follows. In Sec.\ref{secII}, we review the basic framework of TMD evolution for accessing the Sivers asymmetry in the SIDIS process. In Sec.\ref{secIII}, we present the numerical calculation of the asymmetry for the underlying process at the kinematics of HERMES and COMPASS Collaborations, respectively.
The conclusion of the paper is given in Sec.\ref{secIV}.

\section{framework}\label{secII}

In this section we reach the $A_{UT}^{\sin(\phi_h-\phi_S)}$ asymmetry in SIDIS following the TMD factorization procedure in Ref.\cite{Wang:2018pmx}.
We consider the single hadron production in SIDIS by exchanging a virtual photon $q_\mu = l_\mu-l_\mu'$ with an invariant mass $Q^2=-q^2$,
\begin{eqnarray} \label{eq1}
\begin{aligned}
e(l)+p(P) \to e(l')+h(P_h)+X,
\end{aligned}
\end{eqnarray}
where a lepton scatters off a target nucleon with a polarization $S$ and momentum $P$. We adopt the usual SIDIS variables \cite{Meng:1991da},
\begin{eqnarray} \label{eq2}
\begin{aligned}
S_{ep}=(l+P)^2, \qquad x_B=\frac{Q^2}{2P \cdot q}, \qquad y=\frac{P \cdot q}{P \cdot l}=\frac{Q^2}{x_B S_{ep}}, \qquad z_h=\frac{P \cdot P_h}{P \cdot q}.
\end{aligned}
\end{eqnarray}
The Sivers asymmetry of the SIDIS process where a unpolarized lepton scattering off a transversely polarized proton, can be defined as
\begin{eqnarray} \label{eq3}
\begin{aligned}
\displaystyle A_{UT}^{\sin(\phi_h-\phi_S)} = \frac{ \frac{d^5 \Delta \sigma}{dx_B dy dz_h d^2 P_{h\perp}}}{\frac{d^5 \sigma}{dx_B dy dz_h d^2 P_{h\perp}}},
\end{aligned}
\end{eqnarray}
where $\frac{d^5 \sigma}{dx_B dy dz_h d^2 P_{h\perp}}$ and $\frac{d^5 \Delta \sigma}{dx_B dy dz_h d^2 P_{h\perp}}$ represent the spin-averaged and spin-dependent differential cross section, respectively.
When $P_{h\perp} \ll Q$, the TMD factorization applies and the differential SIDIS cross section could be written as \cite{Sun:2013hua}
\begin{eqnarray} \label{eq4}
\begin{aligned}
&\frac{d^5 \sigma(S_\perp)}{dx_B dy dz_h d^2 P_{h\perp}} = \sigma_0 [F_{UU}+\varepsilon_{\alpha\beta} S_\perp^\alpha F_{UT}^\beta], 
\end{aligned}
\end{eqnarray}
where 
\begin{eqnarray} \label{eq5}
\begin{aligned}
\sigma_0 = \frac{2\pi\alpha_{em}^2}{Q^2} \frac{1+(1-y)^2}{y}
\end{aligned}
\end{eqnarray}
and $P_{h\perp}$ is the transverse momentum of the final state hadron with respect to the lepton plane. We introduce $\phi_S$ and $\phi_h$ as the azimuthal angles of the proton's transverse polarization vector and the transverse momentum vector of the final-state hadron. 
These angles are defined in the target rest frame with the $\hat{z}$ axis along the virtual-photon momentum and the $\hat{x}$ axis along the lepton transverse momentum, which follows the Trento conventions \cite{Bacchetta:2004jz}. We have only kept the terms we are interested in.

At the low transverse momentum $(P_{h\perp} \ll Q)$ region the structure functions can be expressed in terms of the TMD factorization as \cite{Sun:2013hua,Kang:2015msa}
\begin{eqnarray} \label{eq6}
\begin{aligned}
F_{UU}(Q;P_{h\perp})=\frac{1}{z_h^2} \int \frac{d^2 b}{(2\pi)^2} e^{i\vec{P}_{h\perp} \cdot \vec{b}/z_h} \widetilde{F}_{UU}(Q;b)+Y_{UU}(Q;P_{h\perp})
\\
F_{UT}^\alpha (Q;P_{h\perp})=\frac{1}{z_h^2} \int \frac{d^2 b}{(2\pi)^2} e^{i\vec{P}_{h\perp} \cdot \vec{b}/z_h} \widetilde{F}_{UT}^\alpha (Q;b)+Y_{UT}^\alpha (Q;P_{h\perp}).
\end{aligned}
\end{eqnarray}
In the expressions of both the structure functions, the first term dominates in $P_{h\perp} \ll Q$ region, and the second term dominates in the region of $P_{h\perp} \ge Q$.
Since we focus on the region $P_{h\perp} \ll Q$, where the TMD factorization approximatively applies, we only reserve the $F$ terms and neglect the $Y$ terms.
However, in practice, it is desirable to stress that the contribution of the $Y$ terms might not be negligible in the kinematical regions of the HERMES and part of the COMPASS experiments, where the $Q^2$ of data might not be that large.
This point has been discussed, e.g., in Ref.\cite{Su:2014wpa}.

Therefore, the spin-averaged differential cross section can be written as
\begin{eqnarray} \label{eq7}
\begin{aligned}
\frac{d^5 \sigma}{dx_B dy dz_h d^2 P_{h\perp}} =\frac{1}{z_h^2} \sigma_0 \int \frac{d^2 b}{(2\pi)^2} e^{i\vec{P}_{h\perp} \cdot \vec{b}/z_h} \widetilde{F}_{UU}(Q;b),
\end{aligned}
\end{eqnarray}
and the spin-dependent differential cross section has the form, 
\begin{eqnarray} \label{eq8}%\label{eq0.1}
\begin{aligned}
\sin(\phi_h-\phi_S)\frac{d^5 \Delta \sigma}{dx_B dy dz_h d^2 P_{h\perp}} =\frac{1}{z_h^2} \sigma_0 \varepsilon_{\alpha\beta} S_\perp^\alpha \int \frac{d^2 b}{(2\pi)^2} e^{i\vec{P}_{h\perp} \cdot \vec{b}/z_h} \widetilde{F}_{UT}^\beta (Q;b).
\end{aligned}
\end{eqnarray}
According to the TMD factorization, the structure functions $\widetilde{F}_{UU}$ and $\widetilde{F}_{UT}^\alpha$ can be written as 
\begin{eqnarray} \label{eq9}%\label{eq1} 
\begin{aligned}
&\widetilde{F}_{UU}(Q;b)=H_{UU}(Q;\mu) \sum_q e_q^2 \widetilde{f}_1^q(x_B,b;\zeta_F,\mu)\widetilde{D}_1^{h/q}(z_h,b;\zeta_D,\mu)
\\
&\widetilde{F}_{UT}^\alpha(Q;b)=H_{UT}(Q;\mu)\sum_q e_q^2 \widetilde{f}_{1T}^{\perp \alpha,q(DIS)}(x_B,b;\zeta_F,\mu) \widetilde{D}_1^{h/q}(z_h,b;\zeta_D,\mu)
\end{aligned}
\end{eqnarray}
where $H_{UU}$ and $H_{UT}$ are the hard factors associated with the corresponding hard scatterings. $\zeta_F(\zeta_D)$ is the energy scale acting as a cutoff to regularize the light cone singularity of the TMD distributions. 
$\widetilde{f}_{1}^q$($\widetilde{D}_1^{h/q}$) denotes the subtracted unpolarized distribution (fragmentation) function, and $\widetilde{f}_{1T}^{\perp\alpha,q(DIS)}(x_B,b;\zeta_F,\mu)$ is the subtracted Sivers function in the $b$ space defined as \cite{Wang:2018pmx}
\begin{eqnarray} \label{eq10}
\begin{aligned}
\widetilde{f}_{1T}^{\perp\alpha,q(DIS)}(x_B,b;\zeta_F,\mu)=\int d^2 \vec{k}_\perp e^{-i \vec{k}_\perp \cdot \vec{b}} \frac{k_\perp^\alpha}{M_p} f_{1T}^{\perp,q(DIS)}(x_B,\vec{k}_\perp;\mu).
\end{aligned}
\end{eqnarray}
Here we have the relation $f_{1T}^{\perp,q(DIS)}=-f_{1T}^{\perp,q(DY)}$. 
The hard factors $H_{UU}(Q;\mu)$ and $H_{UT}(Q;\mu)$ are scheme dependent and can be obtained by three different schemes: the Ji-Ma-Yuan scheme \cite{Ji:2004wu}, the CSS scheme \cite{Collins:1981uk,Collins:1984kg}, and the Collins-11 scheme \cite{Collins:2011zzd} in the literature.
It is worth noticing that $H$ factor is absorbed in the CSS formulation into the definition of Wilson coefficient functions, and the final results of the structure function are scheme independent.

\subsection{The unpolarized differential cross section}

From Eq.(\ref{eq9}), we can find that there are two scale parameters named $\zeta_F $(or $\zeta_D$) and $\mu$ in a general TMD PDF. The corresponding evolution equations describe these scale dependences.
The $\zeta$ scale evolution is presented with the Collins-Soper (CS) equation \cite{Collins:1981uk},
\begin{eqnarray} \label{eq11}
\begin{aligned}
\frac{\partial \ln \widetilde{f}_1^q (x_B,b;\zeta_F,\mu)}{\partial \ln \sqrt{\zeta_F}}=\frac{\partial \ln \widetilde{D}_1^{h/q} (z_h,b;\zeta_D,\mu)}{\partial \ln \sqrt{\zeta_D}}=\widetilde{K}(b,\mu),
\end{aligned}
\end{eqnarray}
where $\widetilde{K}(b,\mu)$ denotes the CS kernel.
The $\mu$ dependence originates from  renormalization group equations for $\widetilde{f}_1^q$, $\widetilde{D}_1^{h/q}$, and $\widetilde{K}$
\begin{eqnarray} \label{eq12}
\begin{aligned}
\frac{d\widetilde{K}(b,\mu)}{d\ln\mu}&=-\gamma_K(\alpha_s(\mu))
\\
\frac{d\ln\widetilde{f}_1^q(x_B,b;\zeta_F,\mu)}{d\ln\mu}&=\gamma_F (\alpha_s(\mu),\zeta_F/\mu^2)
\\
\frac{d\ln\widetilde{D}_1^{h/q}(z_h,b;\zeta_D,\mu)}{d\ln\mu}&=\gamma_D (\alpha_s(\mu),\zeta_D/\mu^2),
\end{aligned}
\end{eqnarray}
where $\gamma_K$, $\gamma_F$, and $\gamma_D$ are anomalous dimensions of $\widetilde{K}$, $\widetilde{f}_1^q$, and $\widetilde{D}_1^{h/q}$, respectively.
On the grounds of many of the previous discussions on the solutions of the above equations in Refs.\cite{Collins:1981uk,Collins:1984kg,Collins:2011zzd,Ji:2004wu,Collins:2014jpa,Ji:2004xq}, for a numerical calculation we have to make a choice for the values of $\zeta_F$ and $\zeta_D$. As stated in Ref.\cite{Aybat:2011zv}, we will treat the PDFs and FFs symmetrically and use $\sqrt{\zeta_F}=\sqrt{\zeta_D}=Q$. Then, we can express $\widetilde{f}(x,b;\zeta_F=Q^2,\mu=Q)$ as $\widetilde{f}(x,b,Q)$ for simplicity.
Therefore, we can summarize that the energy evolution of TMDs $(\widetilde{f})$ from an initial energy $\mu$ to another energy $Q$ can be represented by the Sudakov form factor in the exponential form $\exp(-S)$,
\begin{eqnarray} \label{eq13}
\begin{aligned}
\widetilde{f}(x,b,Q)=\mathcal{F} \cdot e^{-S} \cdot \widetilde{f}(x,b,\mu),
\end{aligned}
\end{eqnarray}
where $\mathcal{F}$ is the hard factor depending on the scheme one chooses. 

We consider the evolution of a TMD function $\widetilde{f}(x,\vec{k}_\perp;Q)$ probed at an energy scale $Q$ carrying a collinear momentum fraction $x$ and a transverse momentum $\vec{k}_\perp$.
It is convenient to reach an energy evolution in the coordinate space; thus, we adopt the Fourier transform of $\widetilde{f}(x,\vec{k}_\perp;Q)$ in the two-dimensional $b$ space listed as
\begin{eqnarray} \label{eq14}
\begin{aligned}
\widetilde{f}(x,b;Q)=\int d^2 \vec{k}_\perp e^{-i\vec{k}_\perp \cdot \vec{b}} \widetilde{f}(x,\vec{k}_\perp;Q).
\end{aligned}
\end{eqnarray}

In this paper, we employ the Collins-Soper-Sterman(CSS) formalsim and pick an initial scale $Q_i=c/b$ for energy evolution. Here, $c=2e^{-\gamma_E}$, and $\gamma_E \approx 0.577$ is the Euler's constant. The energy evolution of TMD in the $b$ space from an initial scale $Q_i$ up to the scale $Q_f=Q$ is represented by \cite{Collins:2011zzd,Aybat:2011zv,Aybat:2011ge,Echevarria:2012pw}
\begin{eqnarray} \label{eq15}
\begin{aligned}
\widetilde{f}(x,b;Q)=\widetilde{f}(x,b;c/b)\exp\left\{ -\int_{c/b}^Q \frac{d\mu}{\mu} \left( A\ln\frac{Q^2}{\mu^2}+B \right) \right\}\left( \frac{Q^2}{(c/b)^2} \right)^{-D}.
\end{aligned}
\end{eqnarray}
The coefficients $A$, $B$ and $D$ can be expanded as a $\alpha_s/\pi$ series,
\begin{eqnarray} \label{eq16}
\begin{aligned}
&A=\sum_{n=1}^\infty A^{(n)}\left(\frac{\alpha_s}{\pi}\right)^n
\\
&B=\sum_{n=1}^\infty B^{(n)}\left(\frac{\alpha_s}{\pi}\right)^n
\\
&D=\sum_{n=1}^\infty D^{(n)}\left(\frac{\alpha_s}{\pi}\right)^n.
\end{aligned}
\end{eqnarray}
In our calculation, we will take $A^{(1)}$, $A^{(2)}$, and $B^{(1)}$ up to the NLL accuracy,
\begin{eqnarray} \label{eq17}
\begin{aligned}
&A^{(1)}=C_F
\\
&A^{(2)}=\frac{C_F}{2}\left[ C_A \left(\frac{67}{18}-\frac{\pi^2}{6}\right)-\frac{10}{9} T_R n_f \right]
\\
&B^{(1)}=-\frac{3}{2}C_F
\\
&D^{(1)}=0,
\end{aligned}
\end{eqnarray}
where $C_F=\frac{4}{3}$, $C_A=3$, and $T_R=\frac{1}{2}$ are color factors,
$n_f=5$ is the the quark-antiquark active number of flavors into which the gluon may split.
By Fourier transforming back in the transverse momentum space, one obtains
\begin{eqnarray} \label{eq18}
\begin{aligned}
\widetilde{f}(x,k_\perp;Q)=\int \frac{d^2 b}{(2\pi)^2} e^{i\vec{k}_\perp \cdot \vec{b}} \widetilde{f}(x,b;Q)=\frac{1}{2\pi}\int_0^\infty db b J_0(k_\perp b) \widetilde{f}(x,b;Q)
\end{aligned}
\end{eqnarray}
where $J_0$ is the Bessel function of the zeroth order.
We should handle the details of the whole $b \in [0,\infty]$ region; i.e, we have to extrapolate to the nonperturbative large-$b$ region.  
A nonperturbative Sudakov factor $R_{\rm NP}(x,b;Q)=\exp(-S_{\rm NP})$ is introduced by
\begin{eqnarray} \label{eq19}
\begin{aligned}
\widetilde{f}(x,b;Q)=\widetilde{f}_{\rm pert}(x,b_*;Q) R_{\rm NP}(x,b;Q),
\end{aligned}
\end{eqnarray}
where the perturbative part of the TMD $\widetilde{f}(x,b_*;Q)$ comes to be
\begin{eqnarray} \label{eq20}
\begin{aligned}
\widetilde{f}_{\rm pert}(x,b_*;Q)=\widetilde{f}\left(x,b;\frac{c}{b_*}\right)S_{\rm pert}(Q;b),
\end{aligned}
\end{eqnarray}
which is valid only when $1/b \ll \Lambda_{QCD}$ and $b_*=\frac{b}{\sqrt{1+(b/b_{\rm max})^2}}$. It has the property that $b_* \approx b$ at low values of $b$ and $b_* \approx b_{\rm max}$ at the large $b$ values.
The typical value of $b_{\rm max}$ is chosen about 1 GeV$^{-1}$ so that $b_*$ is always in the perturbative region.
This $b_*$-prescription introduces a cutoff value $b_{\rm max}$ and allows for a smooth transition from the perturbative region and avoids the Landau pole singularity in $\alpha_s$.
Then the total Sudakov-like form factor can be written as the sum of the perturbatively calculable part and the nonperturbative contribution
\begin{eqnarray} \label{eq21}
\begin{aligned}
S(Q;b)=S_{\rm pert}(Q;b_*)+S_{\rm NP}(Q;b),
\end{aligned}
\end{eqnarray}
and the perturbative part of the Sudakov form factor can be written as
\begin{eqnarray} \label{eq22}
\begin{aligned}
S_{\rm pert}(Q;b_*)=\int_{c/b^*}^Q \frac{d\mu}{\mu} \left[ A\ln\frac{Q^2}{\mu^2}+B \right].
\end{aligned}
\end{eqnarray}
In the region where $1/b \gg \Lambda_{\rm QCD}$, the TMD PDF(FF) at a fixed scale in $b$ space can be expanded as the convolution of perturbatively calculable hard coefficients and the corresponding collinear PDFs(FFs) \cite{Collins:1981uk,Bacchetta:2013pqa},
\begin{eqnarray} \label{eq23}%\label{eq1.1} 
\begin{aligned}
\widetilde{f}_{q/H}(x,b;\mu)&=\sum_i C_{q \leftarrow i} \otimes f^{i/H}(x,\mu)
\\
\widetilde{D}_{H/q}(z,b;\mu)&=\sum_j \hat{C}_{j \leftarrow q} \otimes D^{H/j}(z,\mu),
\end{aligned}
\end{eqnarray}
where $\otimes$ appears for the convolution in the momentum fraction $x$($z$),
\begin{eqnarray} \label{eq24}
\begin{aligned}
C_{q \leftarrow i} \otimes f^{i/H}(x,\mu) \equiv \int_x^1 \frac{d\xi}{\xi} C_{q \leftarrow i}\left( \frac{x}{\xi},b;\mu,\zeta_F \right)f^{i/H}(\xi,\mu)
\\
\hat{C}_{j \leftarrow q} \otimes D^{H/j}(z,\mu) \equiv \int_{z}^1 \frac{d\xi}{\xi} \hat{C}_{j \leftarrow q}\left( \frac{z}{\xi},b;\mu,\zeta_D \right)D^{H/j}(\xi,\mu).
\end{aligned}
\end{eqnarray}
Therefore, including the TMD evolution, the considered TMDs can be expressed as 
\begin{eqnarray} \label{eq25}%\label{eq2}
\begin{aligned}
&\widetilde{f}_1^q (x_B,b;Q^2)=e^{-S_{\rm pert}(Q,b_*)-S_{\rm NP}^{f_1}(Q,b)}\widetilde{\mathcal{F}}_q \sum_i C_{q \leftarrow i} \otimes f_1^i(x_B,\mu_b)
\\
&\widetilde{D}_1^q (z_h,b;Q^2)=e^{-S_{\rm pert}(Q,b_*)-S_{\rm NP}^{D_1}(Q,b)}\widetilde{\mathcal{D}}_q \sum_j \hat{C}_{j \leftarrow q} \otimes D_1^{h/j}(z_h,\mu_b).
\end{aligned}
\end{eqnarray}
We adopt the CSS scheme in which the hard factor in Eq.(\ref{eq9}) together with the functions $\widetilde{\mathcal{F}}_q$ and $\widetilde{\mathcal{D}}_q$ are absorbed into the $C$ functions by applying the renormalization group equation for the running coupling constant in these two factors. We can write down $\widetilde{F}_{UU}(b_*)$ as
\begin{eqnarray} \label{eq26}
\begin{aligned}
\widetilde{F}_{UU}(b_*)=\sum_{q,i,j} e_q^2 \left( C_{q \leftarrow i}^{(DIS)} \otimes f_1^i(x_B,\mu_b) \right)\left( \hat{C}_{j \leftarrow q}^{(DIS)} \otimes D_1^{h/j}(z_h,\mu_b) \right),
\end{aligned}
\end{eqnarray}
where $f_1^i(x_B,\mu_b)$ and $D_1^{h/j}(z_h,\mu_b)$ are the usual unpolarized collinear PDF and FF at the scale $\mu_b =c/b_*$ and the $C$ functions become process dependent.
The final expressions for $C^{(DIS)}$ and $\hat{C}^{(DIS)}$ have been reached in the literature, \cite{Nadolsky:1999kb,Koike:2006fn}
\begin{eqnarray} \label{eq27}
\begin{aligned}
&C_{q \leftarrow q'}^{(DIS)}(x,\mu_b)=\delta_{qq'}\left[ \delta(1-x)+\frac{\alpha_s}{\pi}\left( \frac{C_F}{2}(1-x)-2C_F\delta(1-x) \right) \right]
\\
&C_{q \leftarrow g}^{(DIS)}(x,\mu_b)=\frac{\alpha_s}{\pi}T_R x(1-x)
\\
&\hat{C}_{q' \leftarrow q}^{(DIS)}(z,\mu_b)=\delta_{qq'}\left[ \delta(1-z)+\frac{\alpha_s}{\pi}\left( \frac{C_F}{2}(1-z)-2C_F\delta(1-z)+P_{q \leftarrow q}(z)\ln z \right) \right]
\\
&C_{g \leftarrow q}^{(DIS)}(z,\mu_b)=\frac{\alpha_s}{\pi}\left( \frac{C_F}{2}z + P_{g \leftarrow q}(z)\ln z\right)
\end{aligned}
\end{eqnarray}
with the usual splitting functions $P_{q \leftarrow q}$ and $P_{g \leftarrow q}$ given by
\begin{eqnarray} \label{eq28}
\begin{aligned}
&P_{q \leftarrow q}(z)=C_F \left[ \frac{1+z^2}{(1-z)_+}+\frac{3}{2}\delta(1-z) \right]
\\
& P_{g \leftarrow q}(z)=C_F \frac{1+(1-z)^2}{z},
\end{aligned}
\end{eqnarray}
where the "+" prescription acts in an integral from $x$ to 1 as (see, e.g., \cite{Koike:2006fn}),
\begin{eqnarray} \label{eq29}
\begin{aligned}
\int_x^1 dy \frac{f(y)}{(1-y)_+}=\int_x^1 dy \frac{f(y)-f(1)}{1-y}+f(1)\ln(1-x) .
\end{aligned}
\end{eqnarray}
Substituting the relations of Eq.(\ref{eq25}) into the factorization formula Eq.(\ref{eq9}), we can write down the $\widetilde{F}_{UU}$ in the $b$ space as
\begin{eqnarray} \label{eq30}
\begin{aligned}
\widetilde{F}_{UU}(Q;b)=e^{-2S_{\rm pert}(Q,b_*)-S_{\rm NP}^{DIS}(Q,b)} \sum_q e_q^2 \left( C_{q \leftarrow i}^{(DIS)} \otimes f_1^i(x_B,\mu_b) \right)\left( \hat{C}_{j \leftarrow q}^{(DIS)} \otimes D_1^{h/j}(z_h,\mu_b) \right)
\end{aligned}
\end{eqnarray}
where the nonperturbative form factor originates from the distribution and fragmentation contributions
\begin{eqnarray} \label{eq31}
\begin{aligned}
S_{\rm NP}^{DIS}(Q,b)=S_{\rm NP}^{f_1}(Q,b)+S_{\rm NP}^{D_1}(Q,b),
\end{aligned}
\end{eqnarray}
and we will follow the parametrization of Ref. \cite{Su:2014wpa},
\begin{eqnarray} \label{eq32}
\begin{aligned}
S_{\rm NP}^{DIS}(Q,b)=\frac{g_1}{2}b^2+g_2 \ln\frac{b}{b_*}\ln\frac{Q}{Q_0}+g_3b^2\left(\frac{x_0}{x_B}\right)^\lambda + \frac{g_h}{z_h^2}b^2
\end{aligned}
\end{eqnarray}
where the initial scale $Q_0^2=2.4$GeV$^2$. The parameters are fitted to the experimental data at this initial scale as $g_1=0.212$, $g_2=0.84$, $g_3=0$, $g_h=0.042$, $x_0=0.01$ and $\lambda=0.2$.
Thus, the spin-averaged differential cross section can be cast into
\begin{eqnarray} \label{eq33}%\label{eq2.1}
\begin{aligned}
\frac{d^5 \sigma}{dx_B dy dz_h d^2 P_{h\perp}} &= \frac{1}{z_h^2} \sigma_0 \int \frac{d^2 b}{(2\pi)^2} e^{i\vec{P}_{h\perp} \cdot \vec{b}/z_h} \widetilde{F}_{UU}(Q;b) \\
&=\frac{1}{z_h^2}\frac{\sigma_0}{2\pi}\int_0^\infty db b J_0\left( \frac{P_{h\perp}b}{z_h} \right) \sum_{q,i,j} C_{q \leftarrow i}^{DIS} \otimes f_1^{i/p}(x_B,\mu_b) \hat{C}_{j \leftarrow q}^{DIS} \otimes D_1^{h/j}(z_h,\mu_b) e^{-2S_{\rm pert}-S_{\rm NP}^{DIS}} .
\end{aligned}
\end{eqnarray}

\subsection{The Sivers differential cross section}

Now we turn to the spin-dependent differential cross section in SIDIS contributed by the Sivers function.
The Sivers function $\widetilde{f}_{1T,q/p}^{\perp \alpha (DIS)}$ can be specified by the convolution of the corresponding $C$ coefficients and the collinear correlation functions as \cite{Kang:2011mr,Su:2014wpa,Wang:2018pmx}
\begin{eqnarray} \label{eq34}
\begin{aligned}
\widetilde{f}_{1T,q/p}^{\perp \alpha (DIS)}(x,b;\mu)=\frac{ib^\alpha}{2}\sum_i \Delta C_{q \leftarrow i}^T \otimes f_{i/p}^{(3)}(x',x'';\mu) .
\end{aligned}
\end{eqnarray}
Here $\Delta C_{q \leftarrow i}^T$ represents the hard coefficients, and $f_{i/p}^{(3)}(x',x'';\mu)$ acts as the twist-three quark-gluon-quark or trigluon correlation function. Assuming that the Qiu-Sterman function $T_{q,F}(x,x)$ is the main contribution of the correlation function, in $b$ space the Sivers function can be expressed as
\begin{eqnarray} \label{eq35}
\begin{aligned}
\widetilde{f}_{1T,q/p}^{\perp \alpha (DIS)}(x,b;Q)=\left( \frac{-ib^\alpha}{2} \right)\widetilde{\mathcal{F}}_{{\rm Siv},q} \sum_{i} \Delta C_{q \leftarrow i}^{T,DIS} \otimes T_{i,F}(x_B,x_B,\mu_b)e^{-S_{\rm pert}-S_{\rm NP}^{\rm Siv}},
\end{aligned}
\end{eqnarray}
where $\widetilde{\mathcal{F}}_{{\rm Siv},q}$ is the factor related to the hard scattering.
The relation between the Qiu-Sterman function $T_{q,F}(x,x)$ and the quark Sivers funtion is given by
\begin{eqnarray} \label{eq36}
\begin{aligned}
T_{q,F}(x,x)=-\int d^2 k_\perp \frac{|k_\perp^2|}{M}f_{1T,q/p}^{\perp DIS}(x,k_\perp)=-2Mf_{1T,q/p}^{\perp (1) DIS}(x) .
\end{aligned}
\end{eqnarray}
Here $
f_{1T,q/p}^{\perp (1) DIS}(x)=-\int d^2 k_\perp \frac{|k_\perp^2|}{2M^2}f_{1T,q/p}^{\perp DIS}(x,k_\perp)$ is the first transverse moment of the Sivers function and $M$ is the mass of the colliding hadron.
Similarly, the $\Delta C$ coefficients are calculated as \cite{Sun:2013hua}
\begin{eqnarray} \label{eq37}
\begin{aligned}
\Delta C_{q \leftarrow q'}^T=\delta_{qq'} \left[ \delta(1-x)+\frac{\alpha_s}{\pi}\left( -\frac{1}{4N_c}(1-x)-2C_F\delta(1-x) \right) \right] .
\end{aligned}
\end{eqnarray}
In addition, we adopt the nonperturbative Sudakov form factor $S_{\rm NP}^{\rm Siv}$ in Ref. \cite{Echevarria:2014xaa} for the Sivers function
\begin{eqnarray} \label{eq38}
\begin{aligned}
S_{\rm NP}^{\rm Siv}(b,Q)=b^2\left( g_1^{\rm Siv}+\frac{g_2}{2}\ln\frac{Q}{Q_0} \right) .
\end{aligned}
\end{eqnarray}
Since the fragmentation part of the Sivers asymmetry is not polarized, the nonperturbative Sudakov form factor for the fragmentation function $S_{\rm NP}^{D_1}$ should be the same as the one in unpolarized cross section case. However, $S_{\rm NP}^{D_1}$ can not be separated from Eq.(\ref{eq32}), which gives the total nonperturbative Sudakov form factor in the unpolarized case. Alternatively, we use the $S_{\rm NP}^{D_1}$ coming from the reference paper \cite{Echevarria:2014xaa} for consistency, which can be parametrized as
\begin{eqnarray} \label{eq39}
\begin{aligned}
S_{\rm NP}^{D_1}(b,Q)=b^2\left( g_1^{\rm ff}+\frac{g_2}{2}\ln\frac{Q}{Q_0} \right) .
\end{aligned}
\end{eqnarray}
The parameters have been obtained as
\begin{eqnarray} \label{eq40}
\begin{aligned}
g_1^{\rm Siv}=0.0705\text{GeV}^2 \qquad g_1^{\rm ff}=\frac{0.0475}{z_h^2}\text{GeV}^2 \qquad g_2=0.16\text{GeV}^2 . 
\end{aligned}
\end{eqnarray}
Since we adopt the Trento convention for angle definitions, which is consistent with the COMPASS experiment \cite{Adolph:2014zba}, the spin dependent differential cross section can be written as
\begin{eqnarray} \label{eq41}
\begin{aligned}
&\sin(\phi_h-\phi_S)\frac{d^5 \Delta \sigma}{dx_B dy dz_h d^2 P_{h\perp}} =\frac{1}{z_h^2} \sigma_0 \varepsilon_{\alpha\beta} S_\perp^\alpha \int \frac{d^2 b}{(2\pi)^2} e^{i\vec{P}_{h\perp} \cdot \vec{b}/z_h} \widetilde{F}_{UT}^\beta (Q;b)
\\
&=\varepsilon_{\alpha\beta} S_\perp^\alpha \frac{1}{z_h^2} \sigma_0 \int \frac{d^2 b}{(2\pi)^2} e^{i\vec{P}_{h\perp} \cdot \vec{b}/z_h} \frac{ib^\beta}{2}\sum_{q,i,j} \Delta C_{q \leftarrow i}^{T,DIS} \otimes T_{i,F}(x_B,x_B,\mu_b) \hat{C}_{j \leftarrow q}^{DIS} \otimes D_1^{h/j}(z_h,\mu_b) e^{-2S_{\rm pert}-S_{\rm NP}^{\rm Siv}-S_{\rm NP}^{D_1}}
\\
&=\sin(\phi_h-\phi_S) \frac{1}{z_h^2} \frac{\sigma_0}{4\pi} \int_0^\infty db b^2 J_1\bigg(\frac{P_{h\perp} b}{z_h}\bigg) \sum_{q,i,j} \Delta C_{q \leftarrow i}^{T,DIS} \otimes T_{i,F}(x_B,x_B,\mu_b) \hat{C}_{j \leftarrow q}^{DIS} \otimes D_1^{h/j}(z_h,\mu_b) e^{-2S_{\rm pert}-S_{\rm NP}^{\rm Siv}-S_{\rm NP}^{D_1}} . 
\end{aligned}
\end{eqnarray}
Thus,
\begin{eqnarray} \label{eq42}
\begin{aligned}
&\frac{d^5 \Delta \sigma}{dx_B dy dz_h d^2 P_{h\perp}} =\frac{1}{z_h^2} \frac{\sigma_0}{4\pi} \int_0^\infty db b^2 J_1\bigg(\frac{P_{h\perp} b}{z_h}\bigg) \sum_{q,i,j} \Delta C_{q \leftarrow i}^{T,DIS} \otimes T_{i,F}(x_B,x_B,\mu_b) \hat{C}_{j \leftarrow q}^{DIS} \otimes D_1^{h/j}(z_h,\mu_b) e^{-2S_{\rm pert}-S_{\rm NP}^{\rm Siv}-S_{\rm NP}^{D_1}}.  
\end{aligned}
\end{eqnarray}

\section{Numerical calculation} \label{secIII}

In this section, we present the numerical results of the $A_{UT}^{\sin(\phi_h-\phi_S)}$ in SIDIS with the unpolarized lepton scattering off the transversely polarized proton at the kinematics of COMPASS  and HERMES experiments, respectively.
In order to obtain the numerical estimate of the denominator in the asymmetry given in Eq.(\ref{eq3}), we employ the NLO set of the CT10 parametrization \cite{Lai:2010vv} for the unpolarized distribution function $f_1(x)$ of the proton.
To get reliable results, we use the NLO fit \cite{deFlorian:2014xna} for the unpolarized parton-to-pion fragmentation function since we apply the TMD evolution at NLL accuracy. Meanwhile, we adopt a recent NLO fit \cite{deFlorian:2017lwf} for the unpolarized parton-to-Kaon fragmentation function. 
For the Sivers differential cross section in SIDIS, we apply the TMD evolution. The CSS evolution of the Qiu-Sterman function has been studied extensively in the literature e.g. \cite{Sun:2013hua,Kang:2008ey,Kang:2012em,Zhou:2008mz}. Following CSS evolution formlism, we have to parametrize the Qiu-Sterman function $T_{q,F}(x,x,\mu)$ in a properly initial scale $\mu$ and then evolve it to the scale $\mu_b=c/b^*$. For this part, we employ a recent parametrization \cite{Echevarria:2014xaa} which assummes that the Qiu-Sterman function is proportional to the usual unpolarized collinear PDFs as
\begin{eqnarray} \label{eq43}
\begin{aligned}
T_{q,F}(x,x,\mu)=N_q\frac{(\alpha_q+\beta_q)^{(\alpha_q+\beta_q)}}{\alpha_q^{\alpha_q} \beta_q^{\beta_q}} x^{\alpha_q} (1-x)^{\beta_q} f_1^q (x,\mu),
\end{aligned}
\end{eqnarray}
where $\mu=2.4$GeV, $N_q$, $\alpha_q$ and $\beta_q$ are given in Table 1 of Ref.\cite{Echevarria:2014xaa}.
Following Ref.\cite{Wang:2018pmx} and Ref.\cite{Kang:2015msa}, where only the homogeneous terms of the evolution kernel are kept in order to reach the evolution of the Qiu-Sterman function and twist-3 fragmentation function $\hat{H}^{(3)}$, respectively.
in this paper, we keep the same. 
This homogeneous term of the Qiu-Sterman function evolution kernel is written as
\begin{eqnarray} \label{eq44}
\begin{aligned}
P_{qq}^{QS} \approx P_{qq}^{f_1}-\frac{N_c}{2}\frac{1+z^2}{1-z}-N_c \delta(1-z),
\end{aligned}
\end{eqnarray}
where $P_{qq}^{f_1}$ is the evolution kernel of the unpolarized PDF and has the same form as the $P_{q \leftarrow q}$ in Eq.(\ref{eq28}).

The numerical solution of Qiu-Sterman function's evolution equation is performed by the QCDNUM evolution package \cite{Botje:2010ay}. The energy evolution of fragmentation function is performed by the built-in timelike evolution in QCDNUM.
The QCD coupling constant using in the evolution package and CSS evolution is
\begin{eqnarray} \label{eq46}
\begin{aligned} 
\alpha_s(Q^2)=\frac{12\pi}{(33-2n_f)\ln(Q^2/\Lambda_{\rm QCD}^2)}\left[ 1-\frac{6(153-19n_f)\ln\ln(Q^2/\Lambda_{\rm QCD}^2)}{(33-2n_f)^2 \ln(Q^2/\Lambda_{\rm QCD}^2)} \right] .
\end{aligned}
\end{eqnarray}
The original code of QCDNUM is modified by us so that the Qiu-Sterman function evolution kernel is added; the initial scale for the evolution is chosen to be $Q_0^2=2.4$GeV$^2$. The QCDNUM code is executed with $\alpha_s(Q_0)=0.327$.

\begin{figure}[htp]
\centering
\includegraphics[scale=0.46]{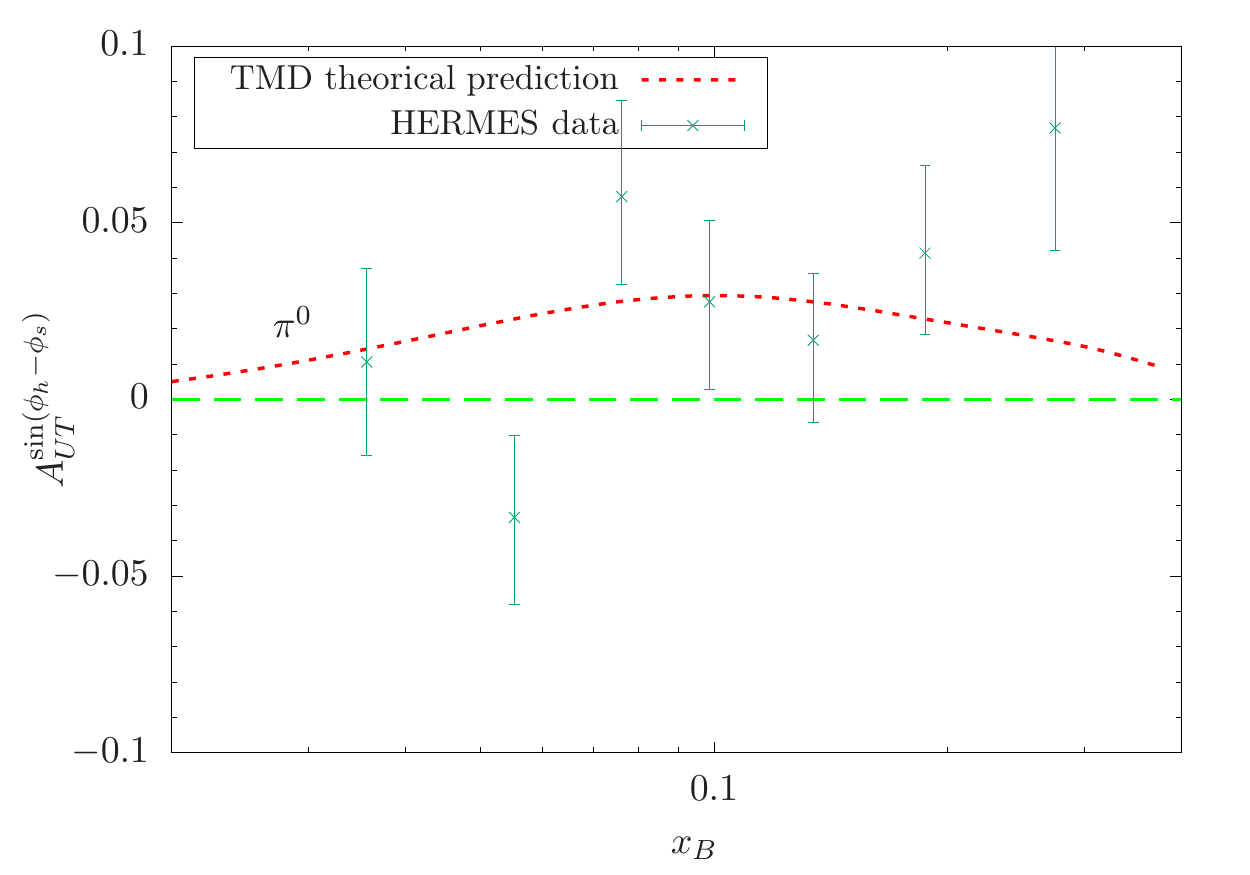}
\includegraphics[scale=0.46]{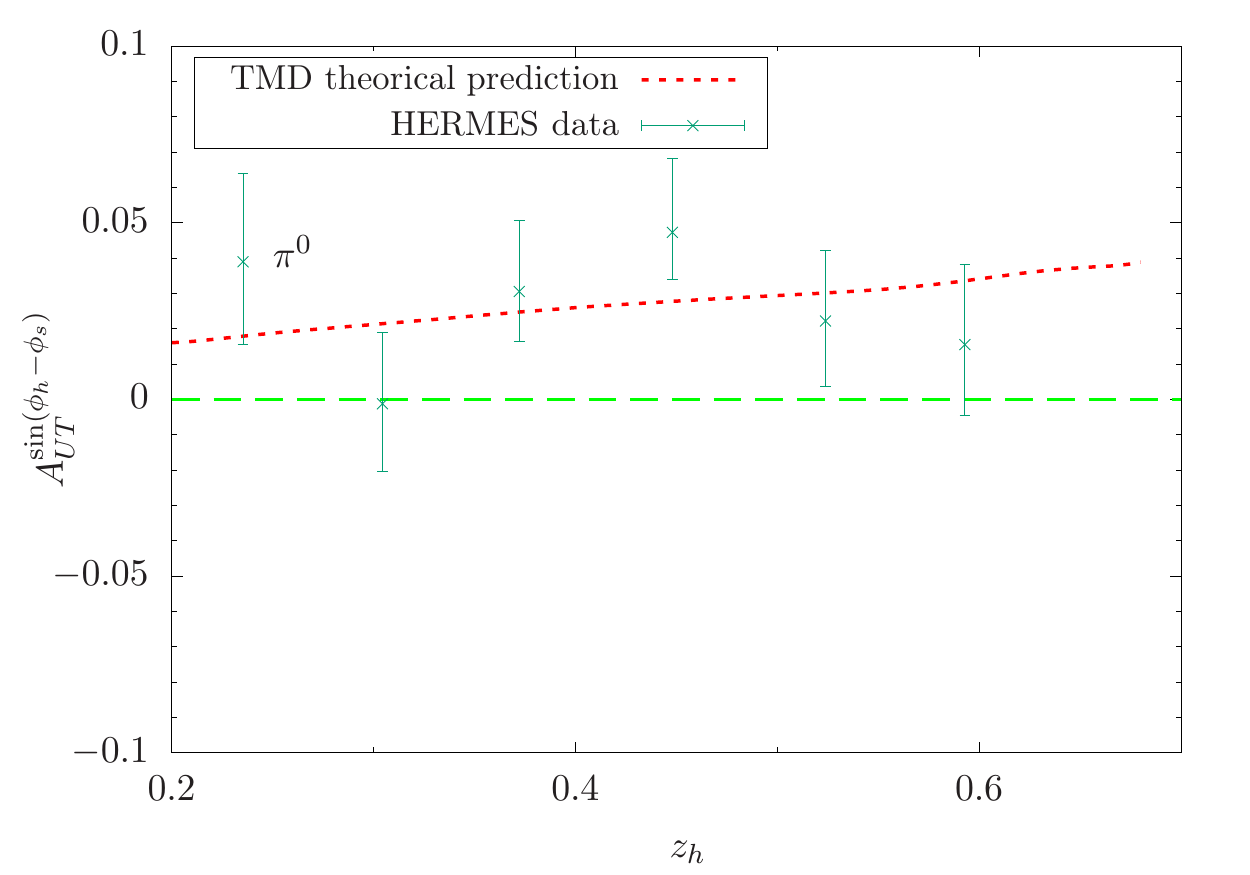}
\includegraphics[scale=0.46]{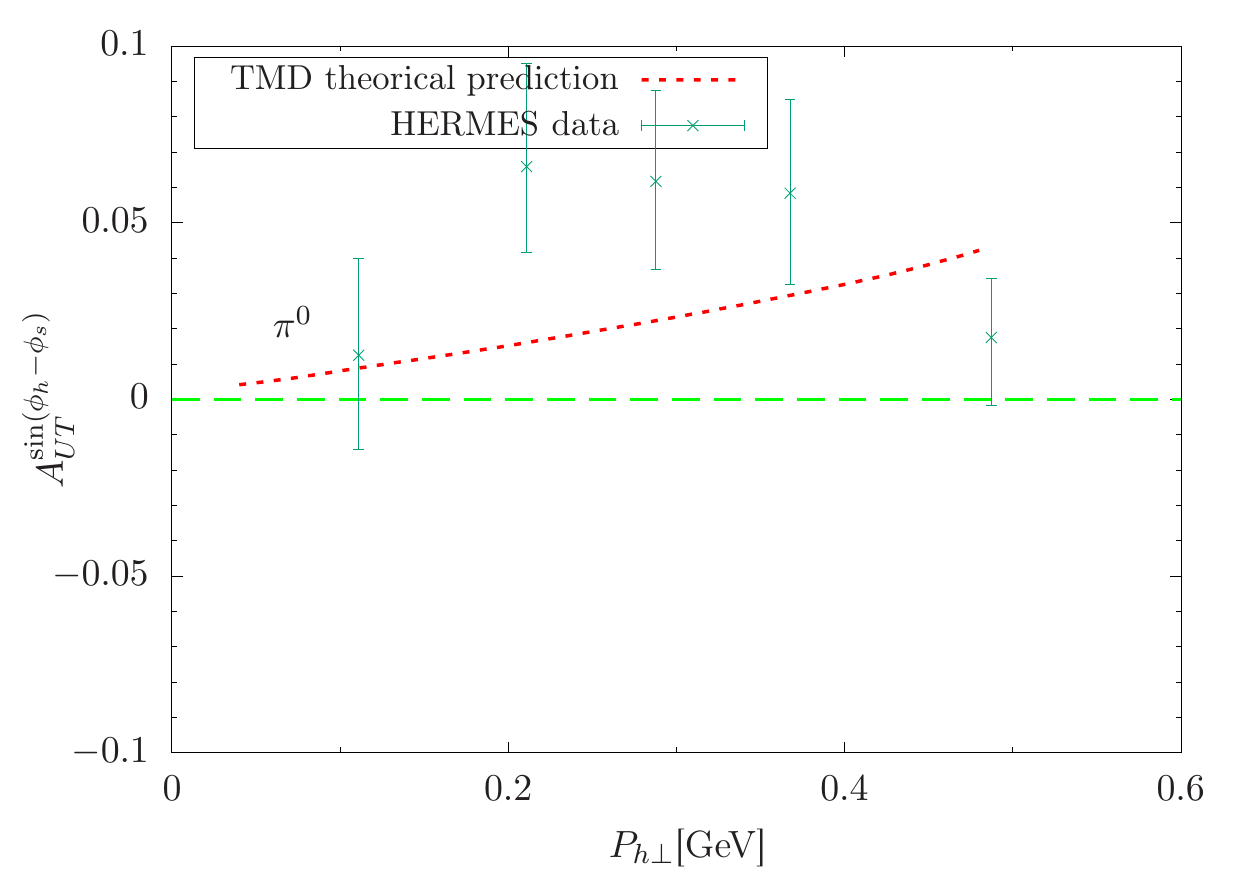}
\caption{ The Sivers asymmetry calculated within TMD factorization, compared with the HERMES measurement \cite{Airapetian:2009ae} for $\pi^0$ production.}
\label{fig2}
\end{figure}
\begin{figure}[htp]
\centering
\includegraphics[scale=0.46]{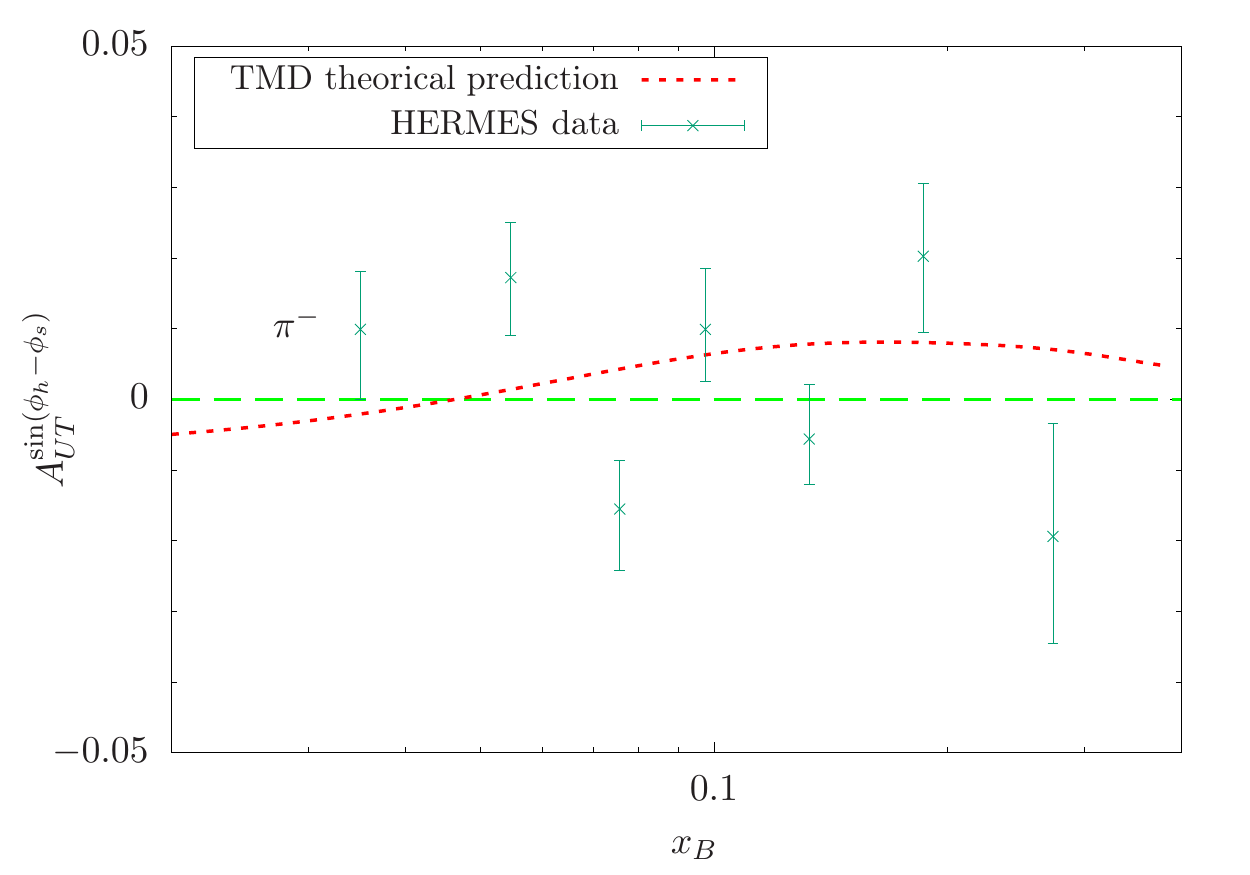}
\includegraphics[scale=0.46]{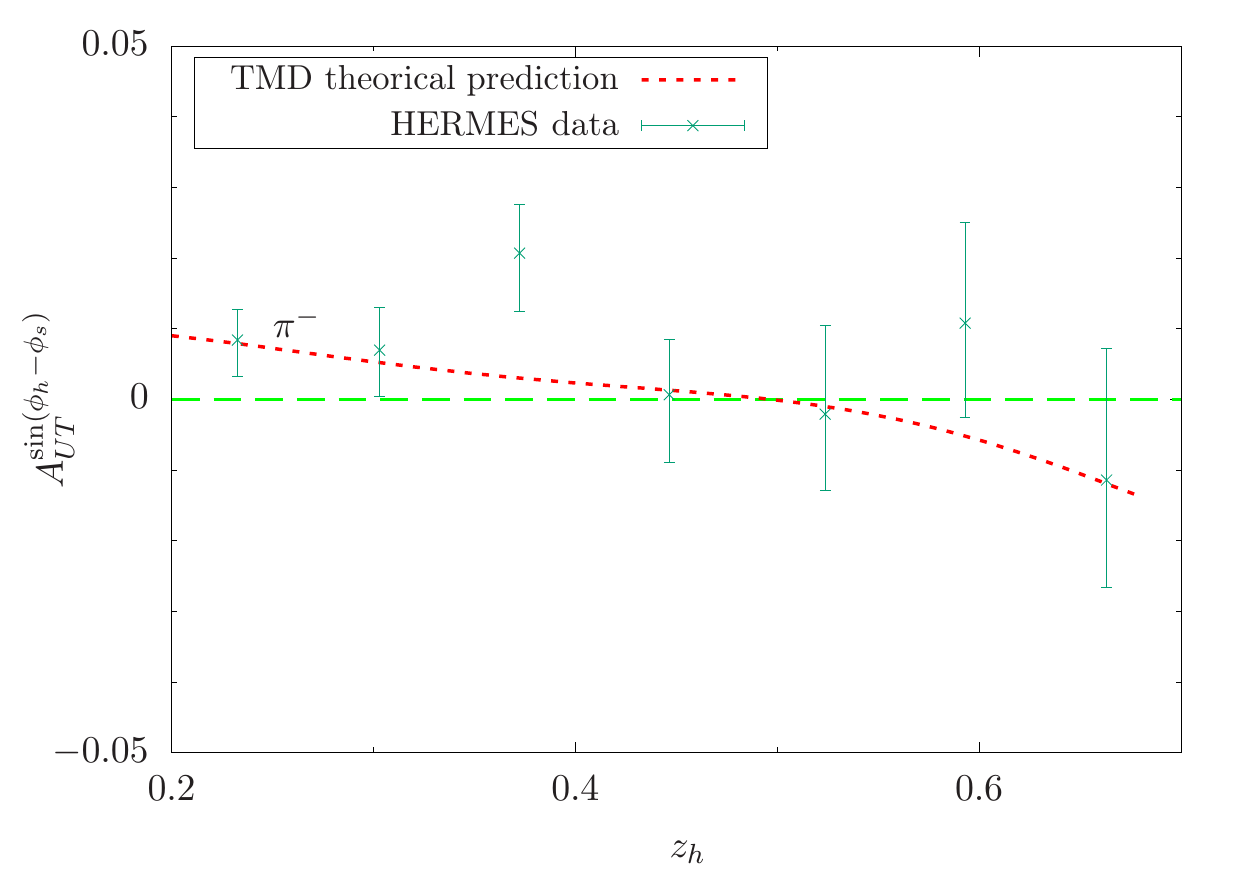}
\includegraphics[scale=0.46]{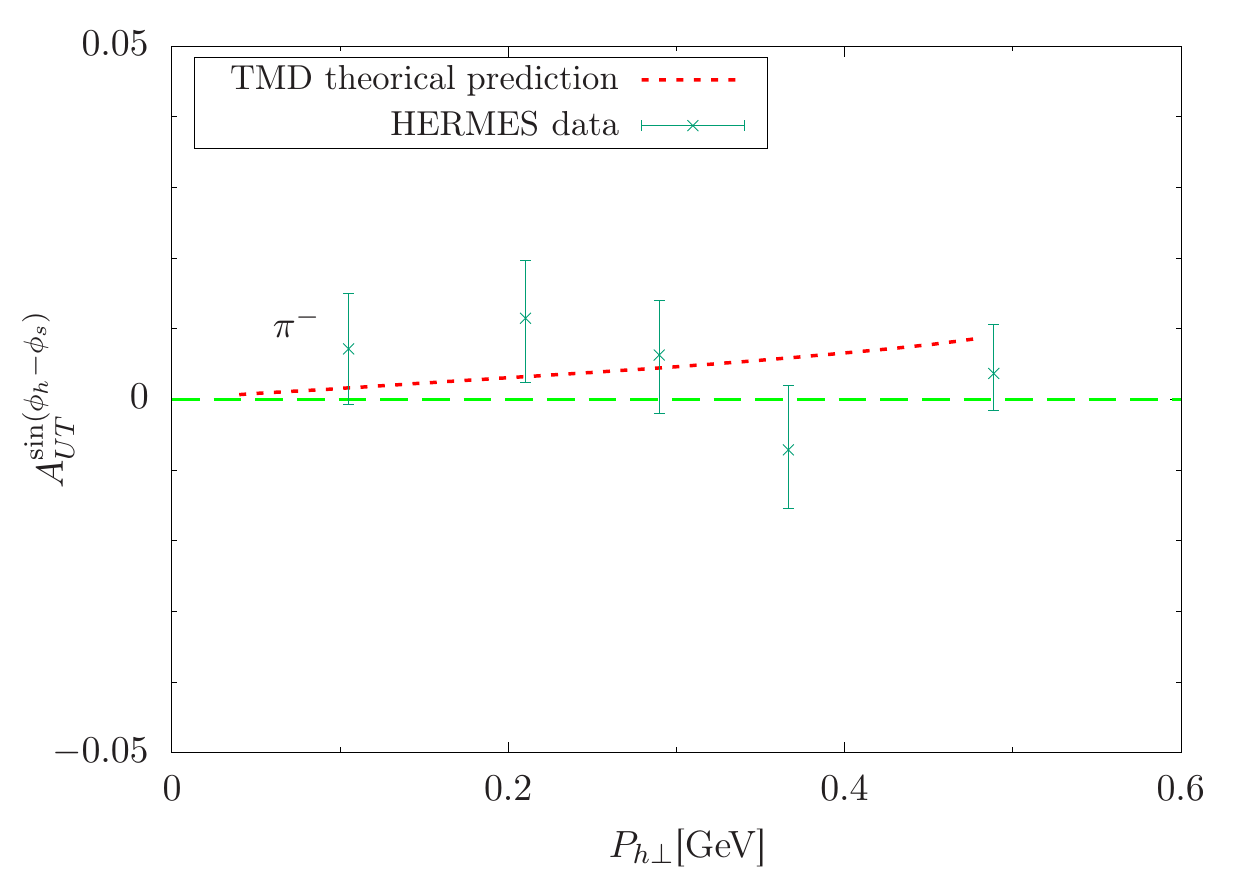}
\caption{ The Sivers asymmetry calculated within TMD factorization, compared with the HERMES measurement \cite{Airapetian:2009ae} for $\pi^-$ production.}
\label{fig3}
\end{figure}
\begin{figure}[htp]
\centering
\includegraphics[scale=0.46]{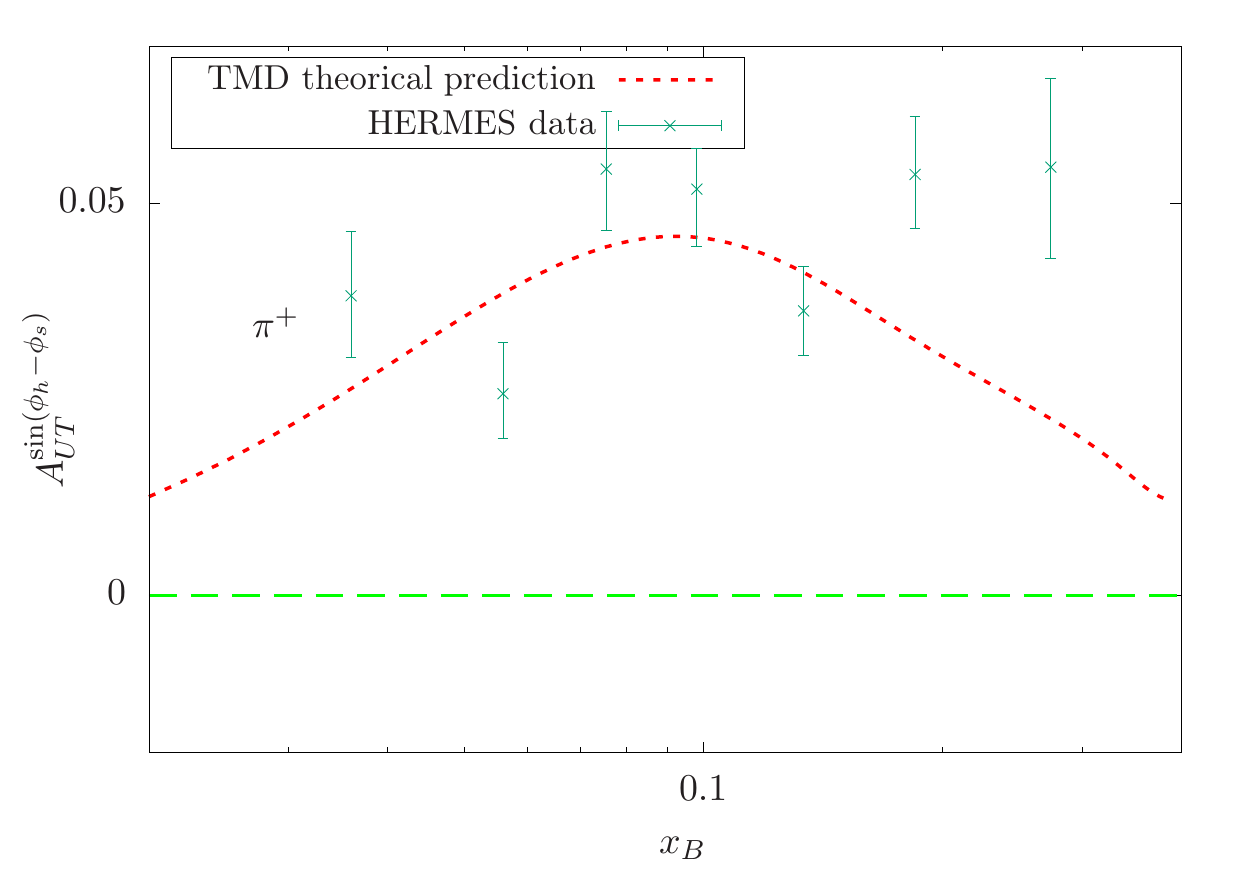}
\includegraphics[scale=0.46]{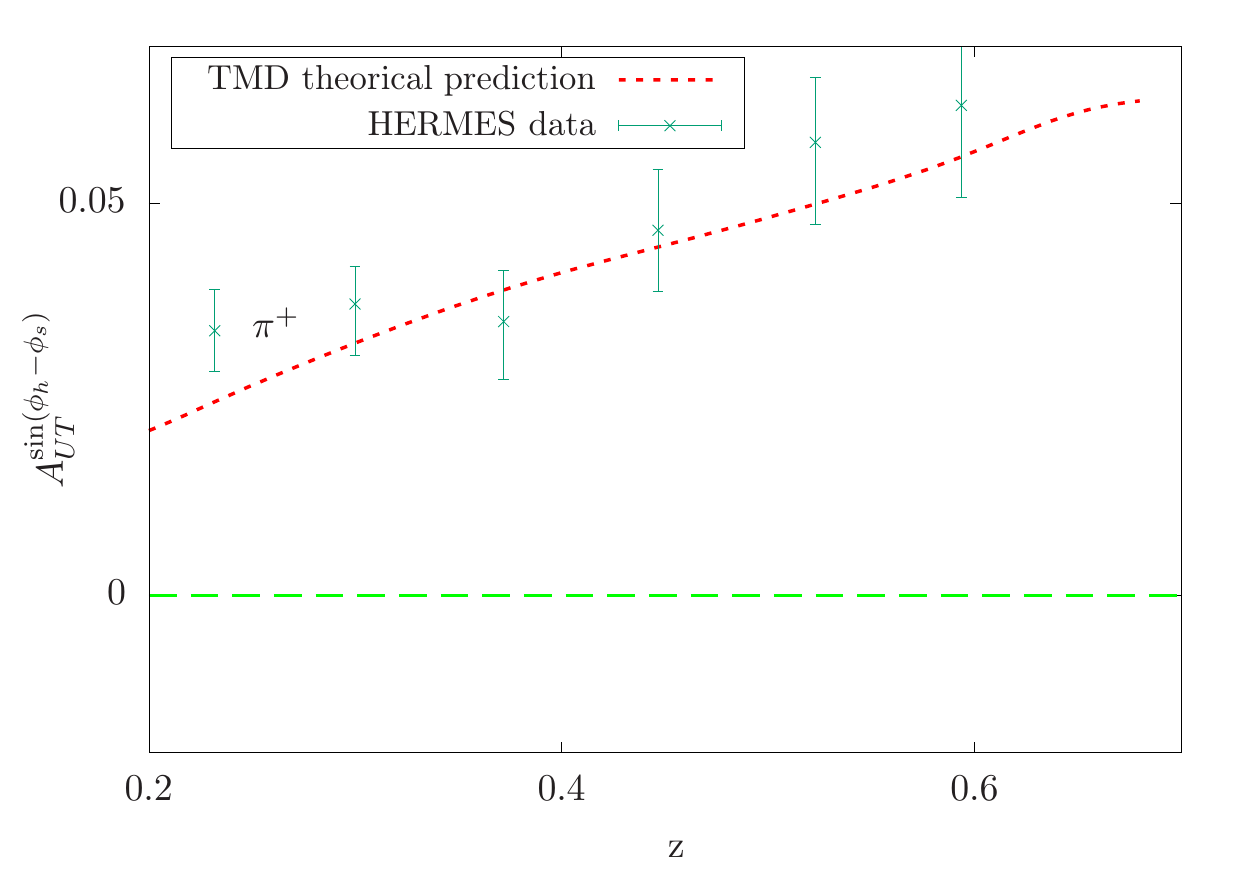}
\includegraphics[scale=0.46]{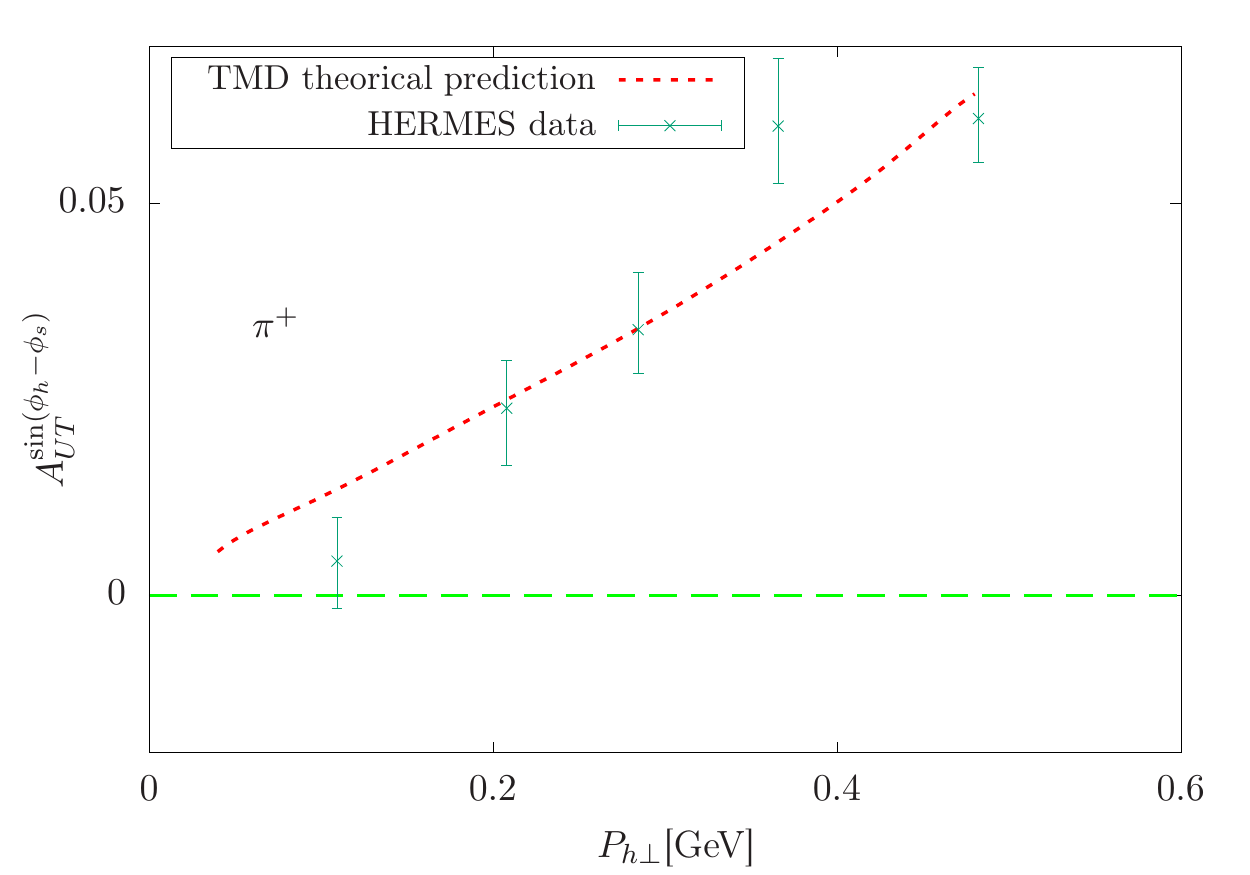}
\caption{ The Sivers asymmetry calculated within TMD factorization, compared with the HERMES measurement \cite{Airapetian:2009ae} for $\pi^+$ production.}
\label{fig4}
\end{figure}
\begin{figure}[htp]
\centering
\includegraphics[scale=0.46]{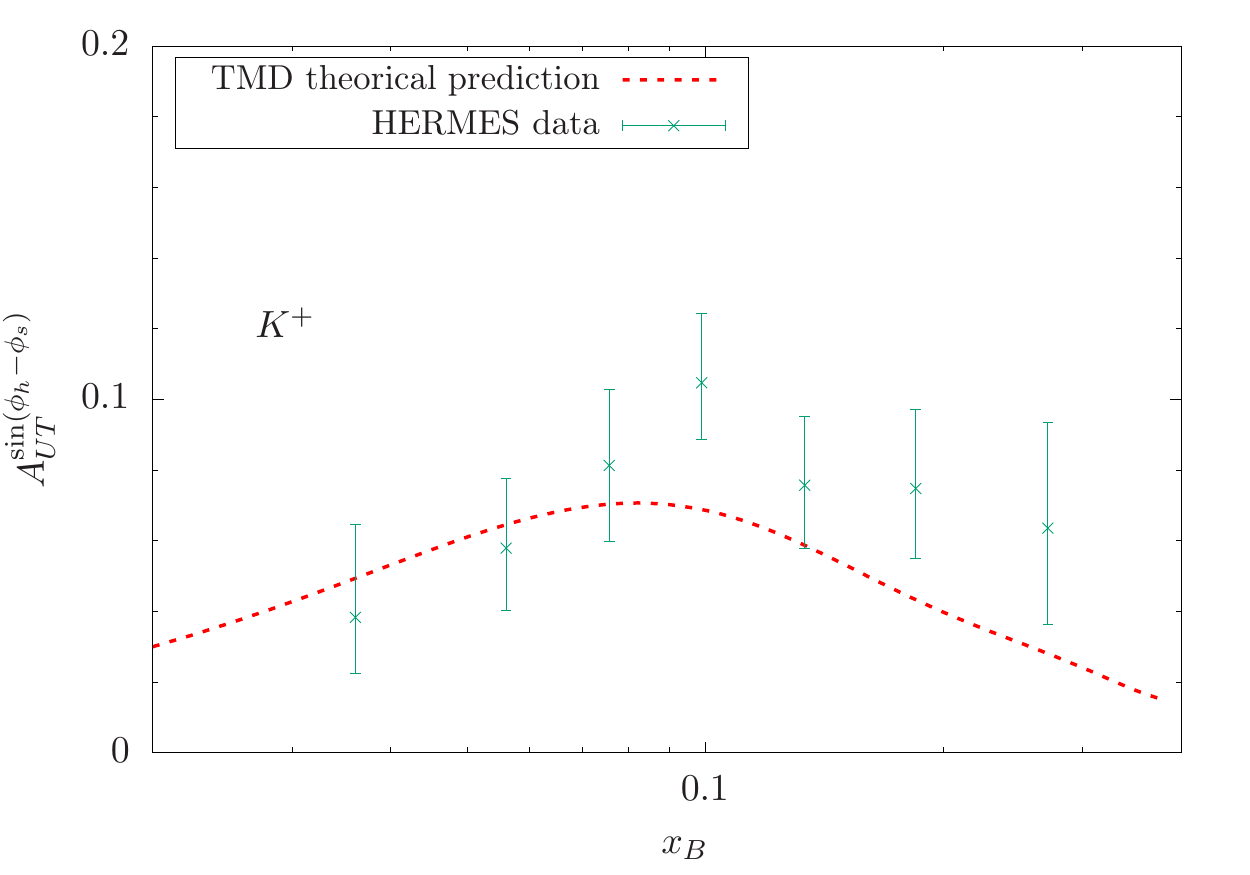}
\includegraphics[scale=0.46]{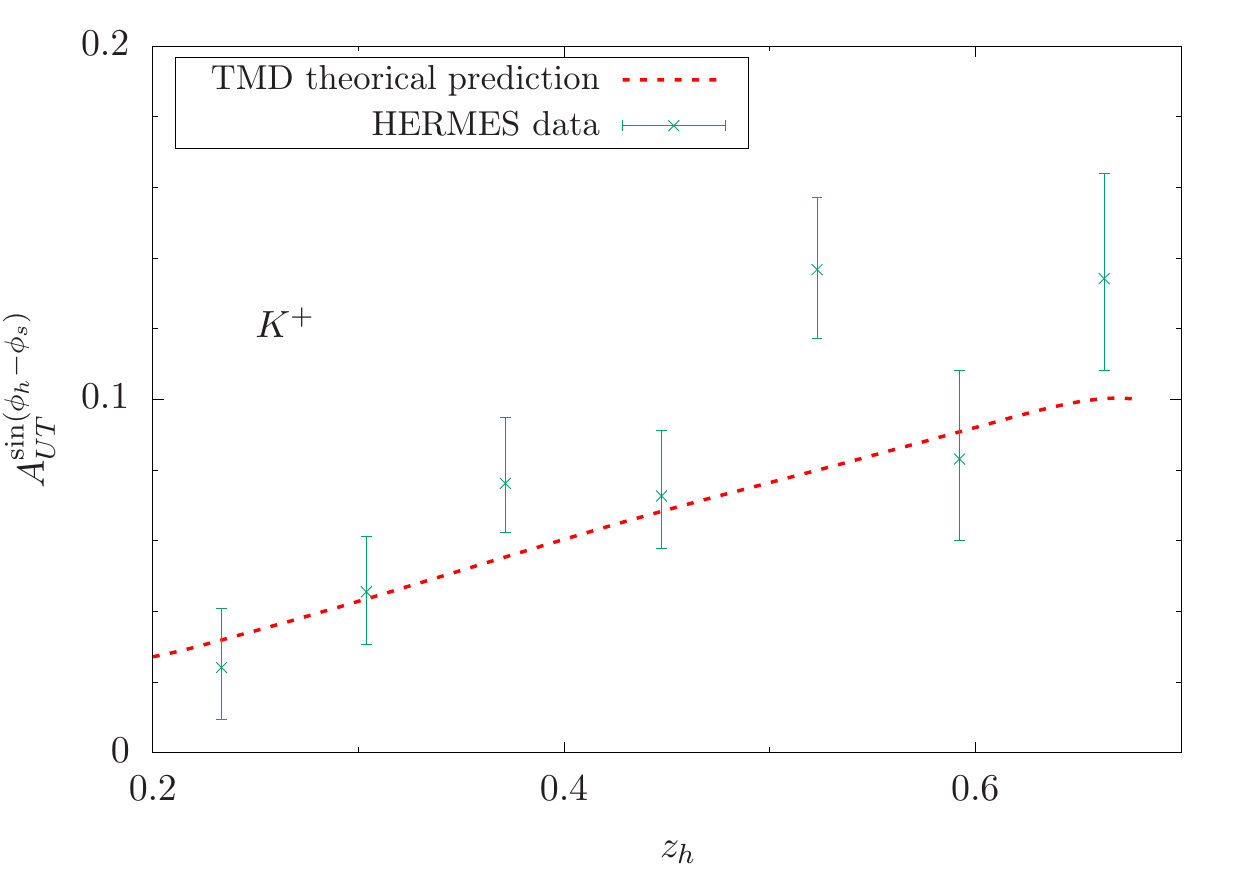}
\includegraphics[scale=0.46]{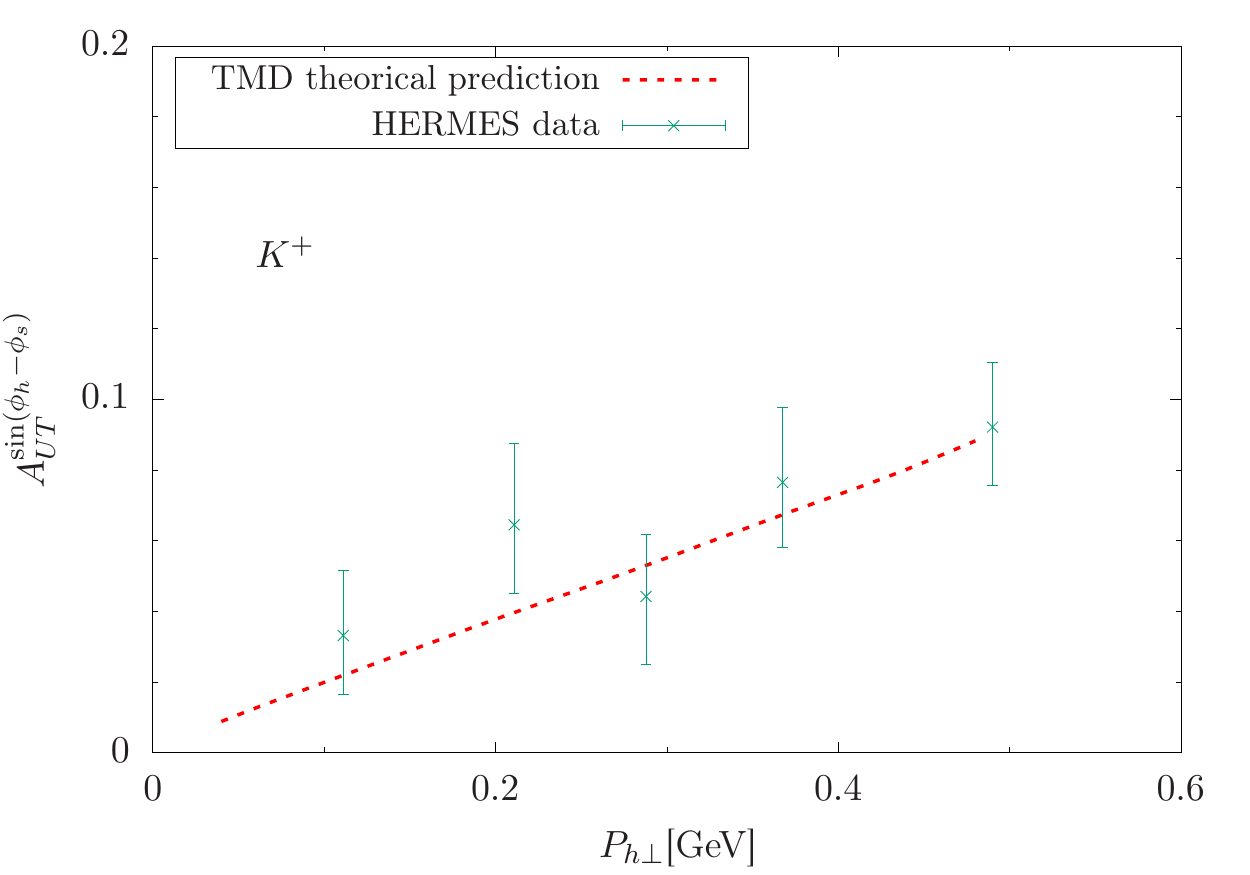}
\caption{ The Sivers asymmetry calculated within TMD factorization, compared with the HERMES measurement \cite{Airapetian:2009ae} for $K^+$ production.}
\label{fig5}
\end{figure}
\begin{figure}[htp]
\centering
\includegraphics[scale=0.46]{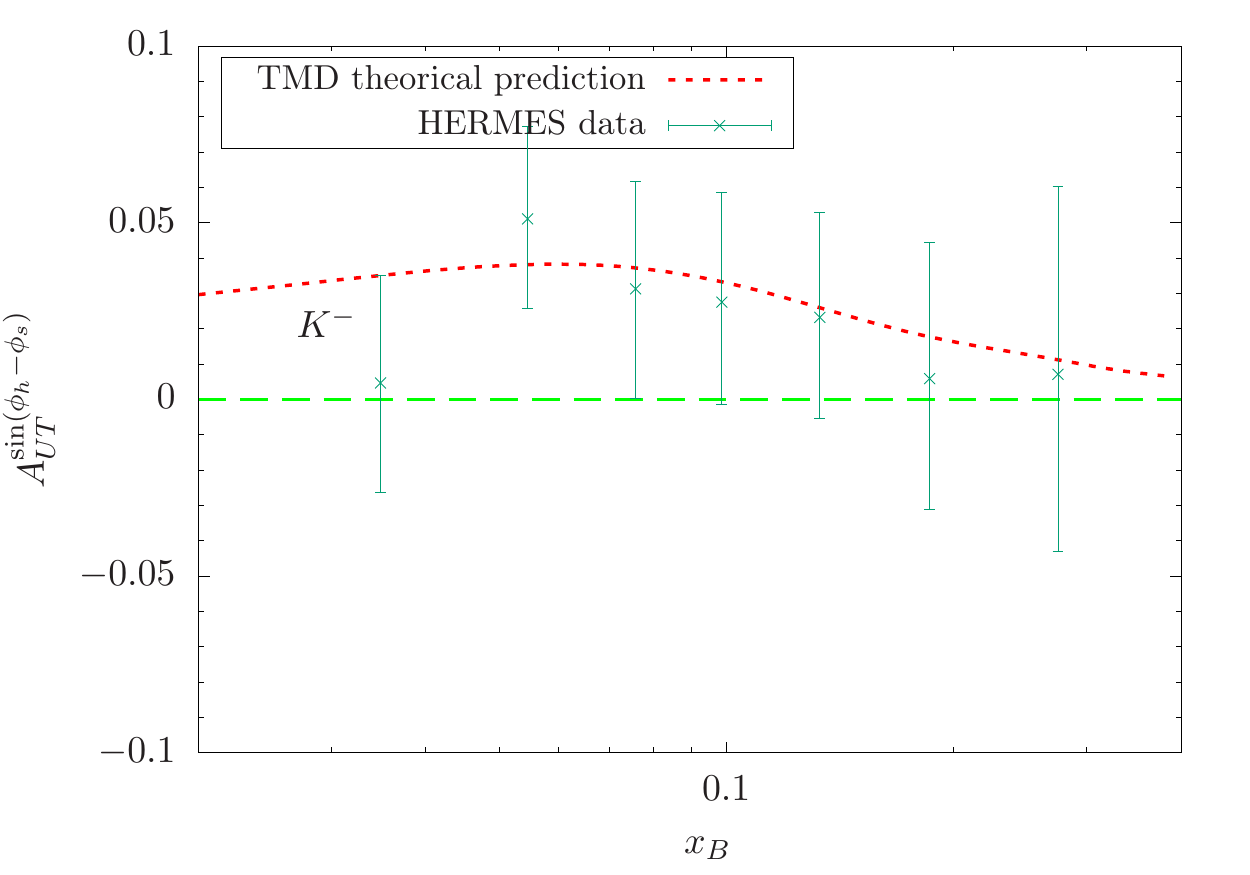}
\includegraphics[scale=0.46]{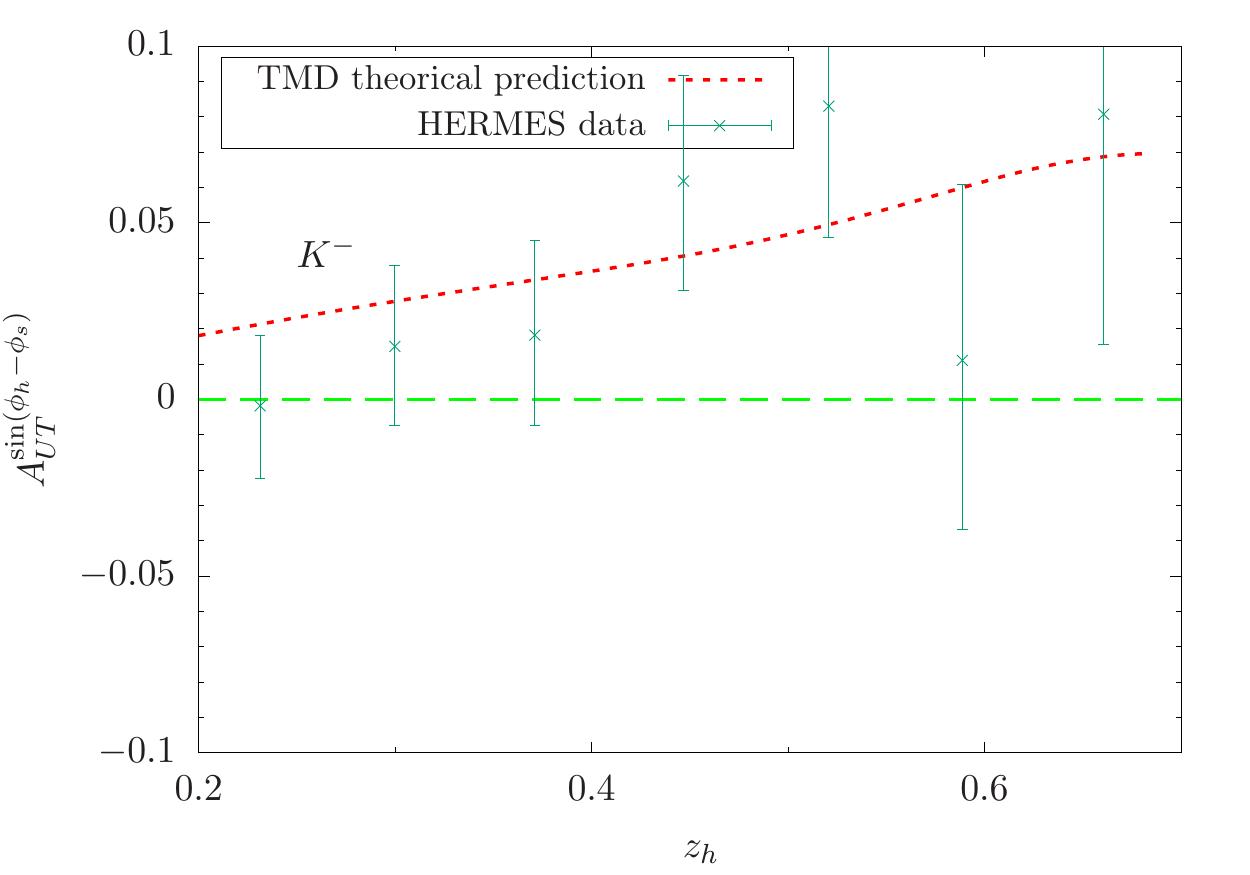}
\includegraphics[scale=0.46]{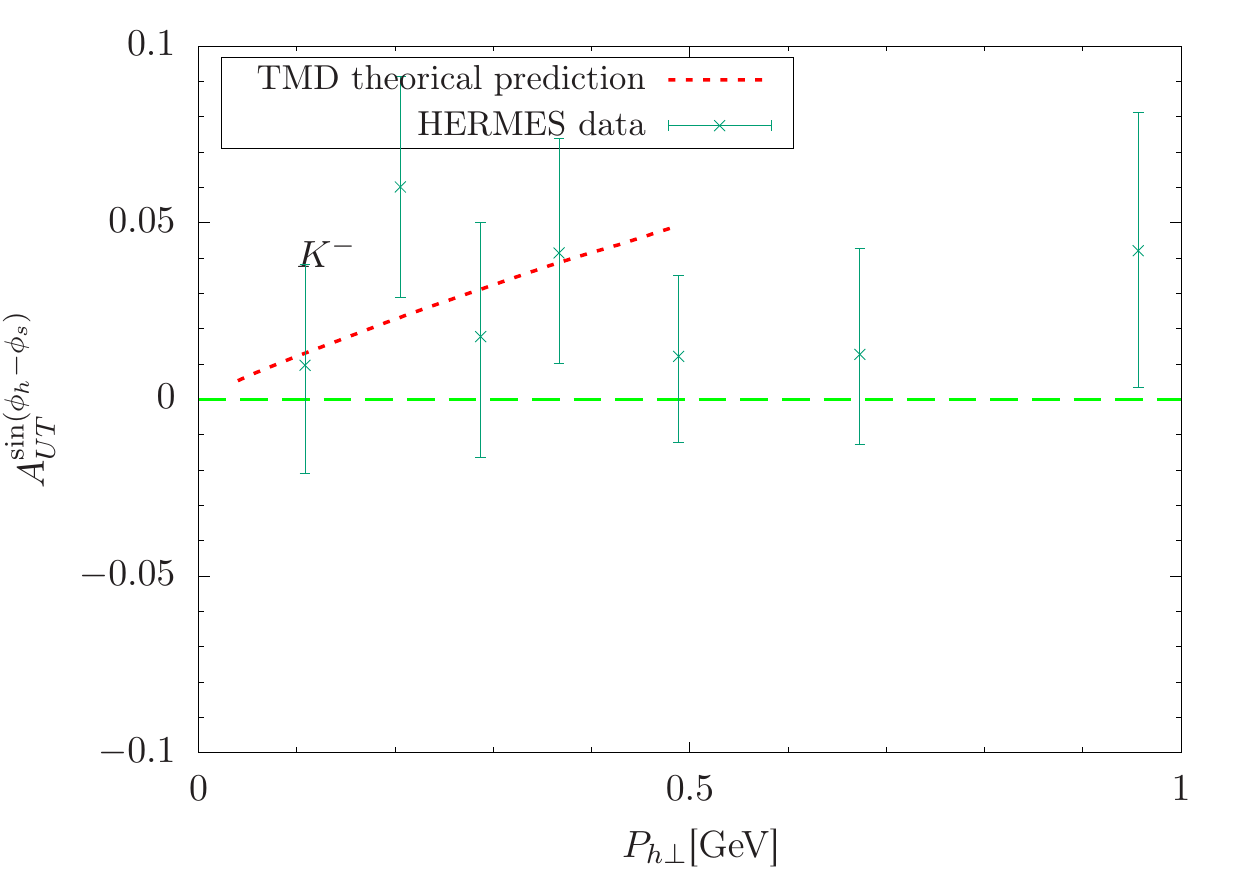}
\caption{ The Sivers asymmetry calculated within TMD factorization, compared with the HERMES measurement \cite{Airapetian:2009ae} for $K^-$ production.}
\label{fig6}
\end{figure}

To perform numerical calculations for $A_{UT}^{\sin(\phi_h-\phi_S)}$ in SIDIS at HERMES, we adopt the following kinematical cuts \cite{Airapetian:2009ae}
\begin{eqnarray} \label{eq47}
\begin{aligned}
&0.023 <x_B< 0.4 \qquad 0.1<y<0.95 \qquad 0.2<z_h<0.7 \qquad P_{h\perp} > 0.1\text{GeV}
\\
&Q^2>1\text{GeV}^2 \qquad W^2>10\text{GeV}^2
\end{aligned}
\end{eqnarray}
where $W$ is the invariant mass of photon-nucleon system with $W^2=(P+q)^2 \approx \frac{1-x_B}{x_B}Q^2$.
Furthermore, like Ref.\cite{Echevarria:2014xaa}, we choose $P_{h\perp} \le 0.5$GeV for hadron production at HERMES since we focus on the region $P_{h\perp} \le Q$ region where the TMD factorization applies.
In Figs.\ref{fig2}-\ref{fig6}, we show the results for pion and kaon production.
The $x_B$-, $z_h$-, and $P_{h\perp}$-dependent asymmetries are depicted in the left, central, and right panels of the figure, respectively. The dashed lines represent our predictions. The full circles with error bars show the preliminary HERMES data for comparison.
For the pion production, Figs.\ref{fig2}-\ref{fig4} give a good description for the HERMES data. The similar conclusion could be reached from Fig. 6 of Ref.\cite{Echevarria:2014xaa} where the authors did not consider the effects of resummation.
Furthermore, the authors parametrize the Qiu-Sterman function by Eq.(\ref{eq43}) for all of the energy scales.
However, as for the kaon (especially $K^-$) production, theoretical results in Fig. 7 of Ref.\cite{Echevarria:2014xaa} underestimate the HERMES data. On the contrary, the theoretical predictions in this paper shown in Fig. \ref{fig5}-\ref{fig6} give a rather good description of the HERMES data, where $x_B$-, $z_h$-, and $P_{h\perp}$-dependent asymmetries are basically distributed within the allowable range of experimental error. In Fig.\ref{fig5}, both the obtained $z_h$- and $P_{h\perp}$-dependent asymmetries for $K^+$ production increase as $z_h$ and $P_{h\perp}$ increase, and the largest asymmetry could arrive at 0.1.
Both the obtained $z_h$- and $P_{h\perp}$-dependent asymmetries for $K^-$ production also increase as $z_h$ and $P_{h\perp}$ increase, and the largest asymmetry could arrive at 0.05.
Then we can reach the conclusion that also after adding the resummation effect and evolving exactly the Qiu-Sterman function using the corresponding evolution kernel, the asymmetry results could be improved to a certain extent.

\begin{figure}[htp]
\centering
\includegraphics[scale=0.46]{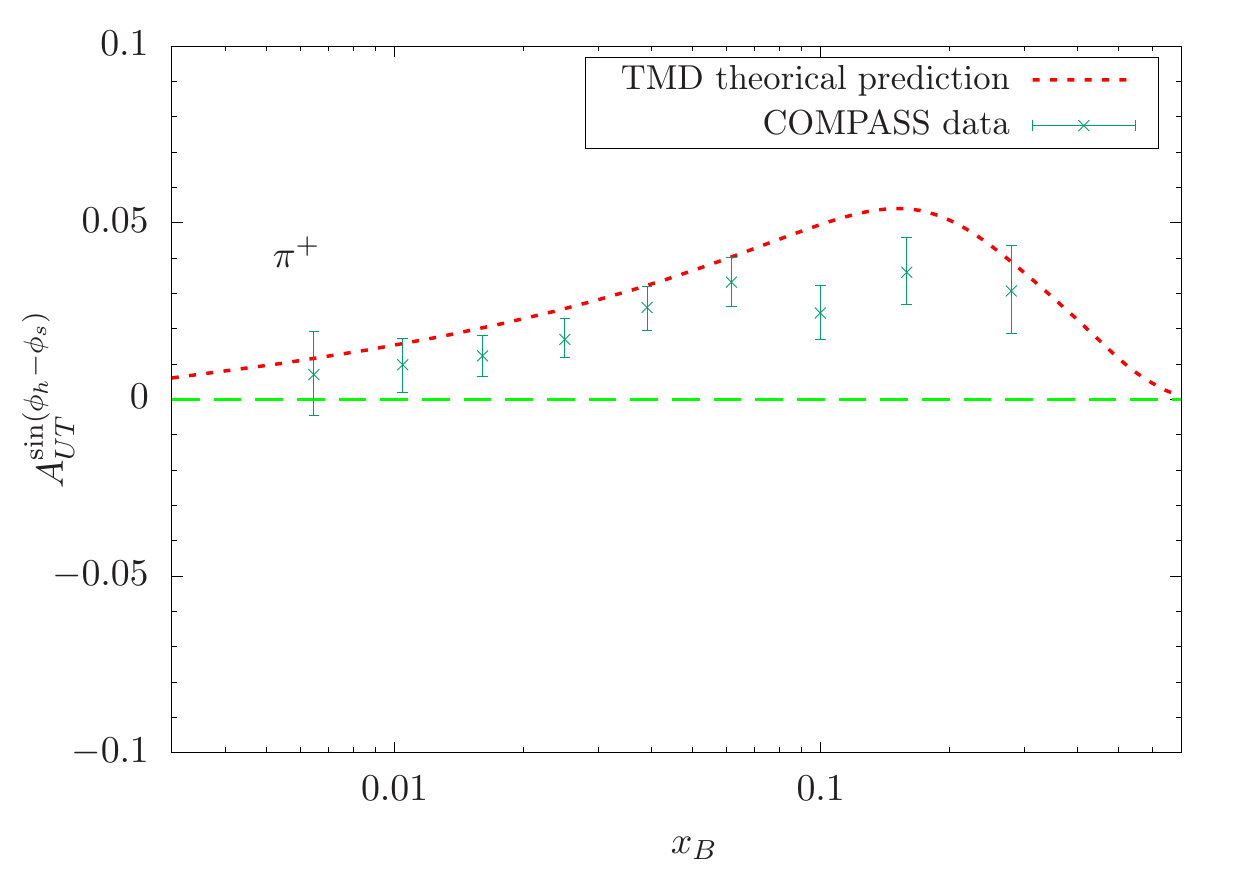}
\includegraphics[scale=0.46]{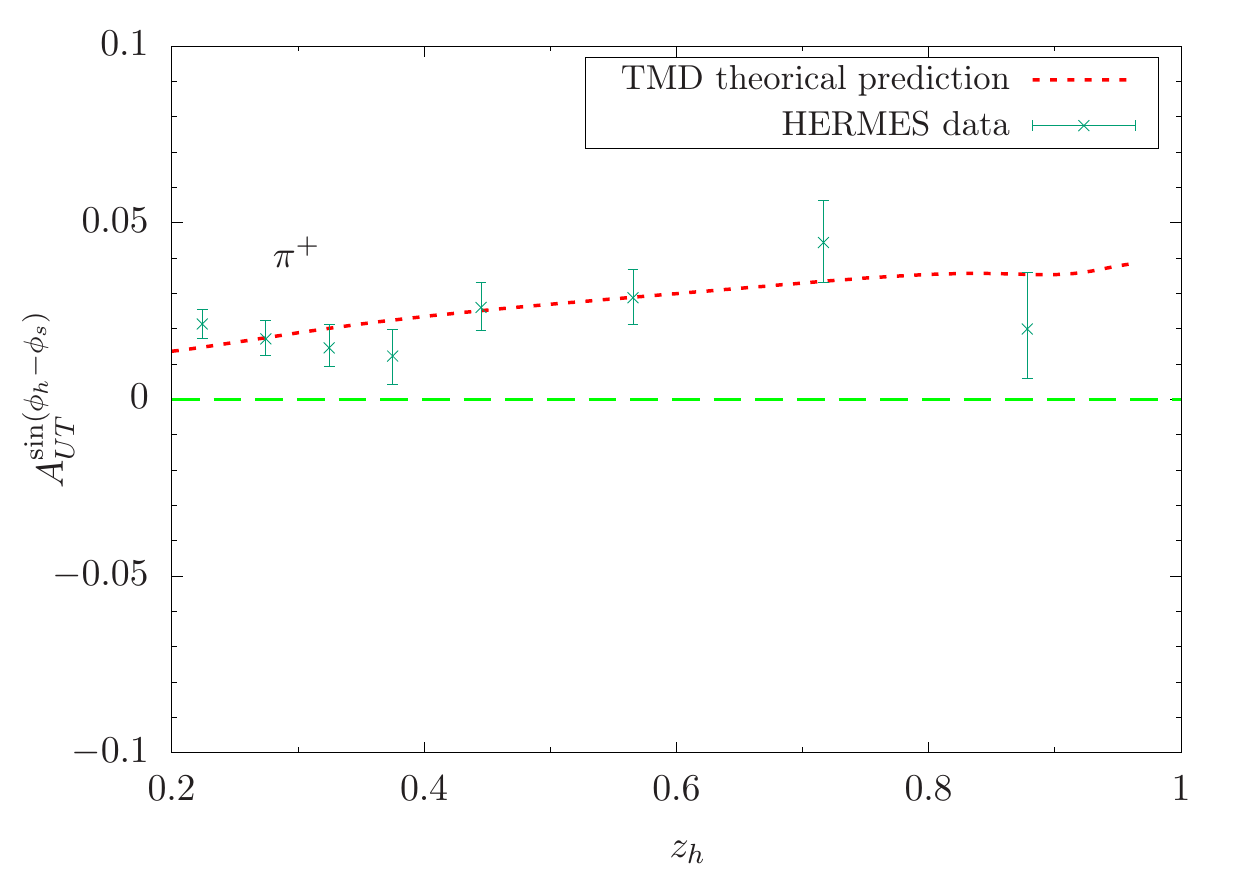}
\includegraphics[scale=0.46]{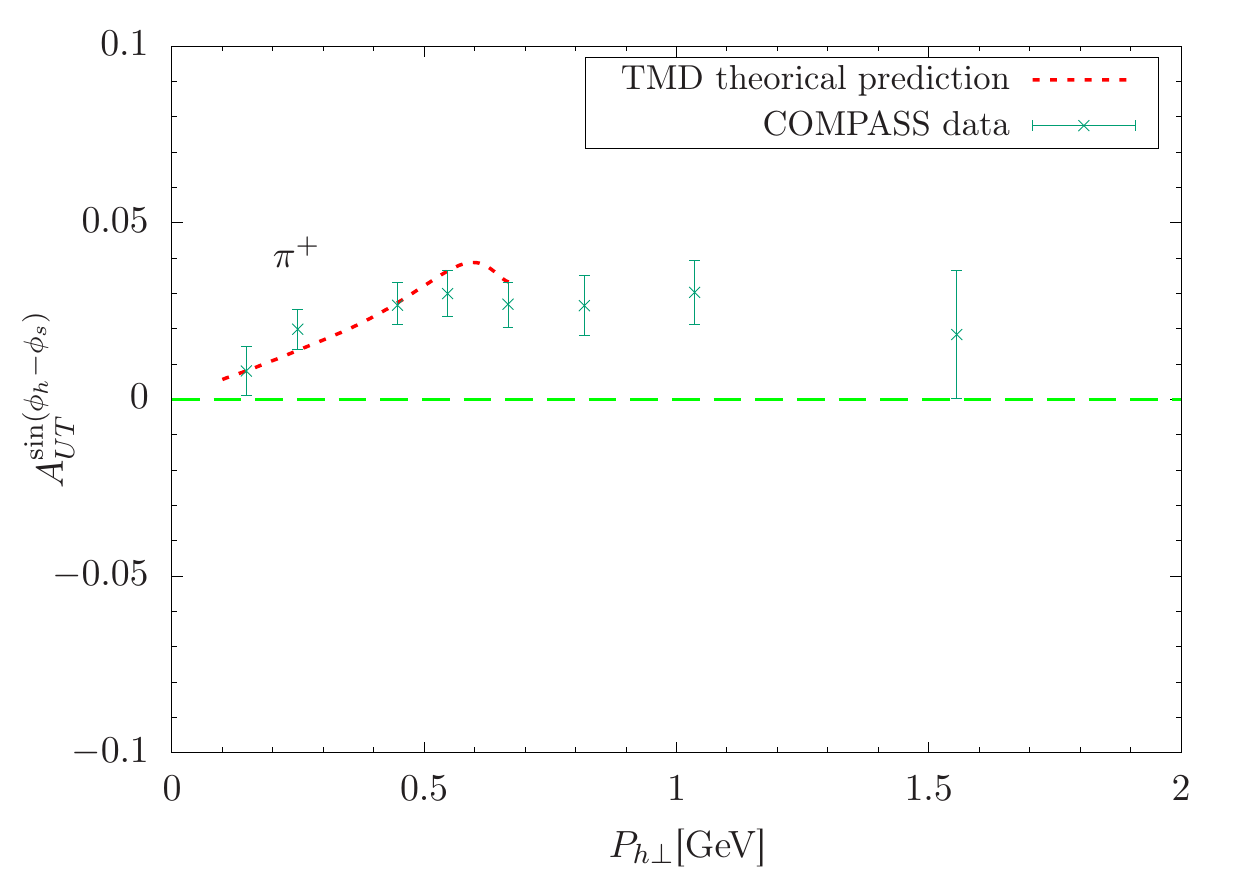}
\caption{ The Sivers asymmetry calculated within TMD factorization, compared with the COMPASS measurement \cite{Adolph:2014zba} for $\pi^+$ production.}
\label{fig7}
\end{figure}
\begin{figure}[htp]
\centering
\includegraphics[scale=0.46]{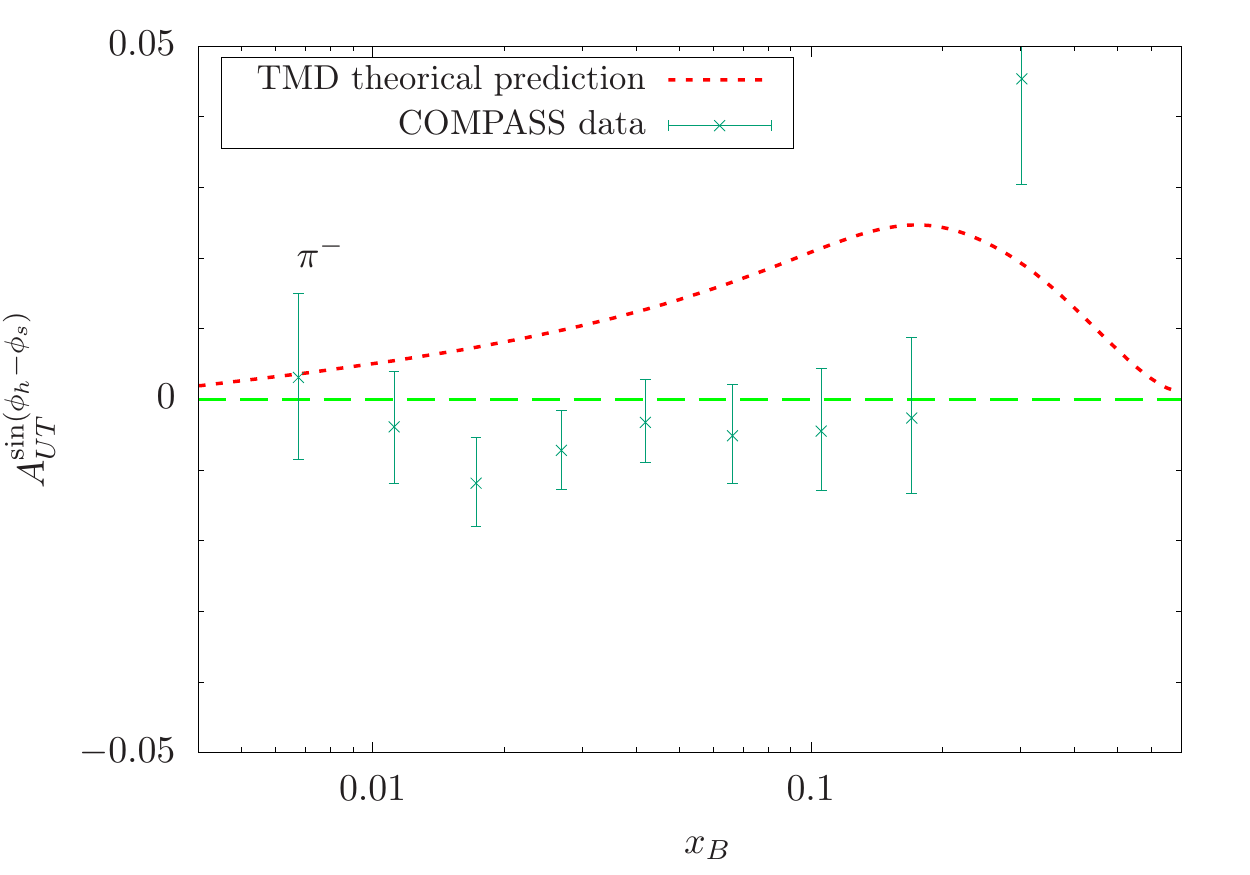}
\includegraphics[scale=0.46]{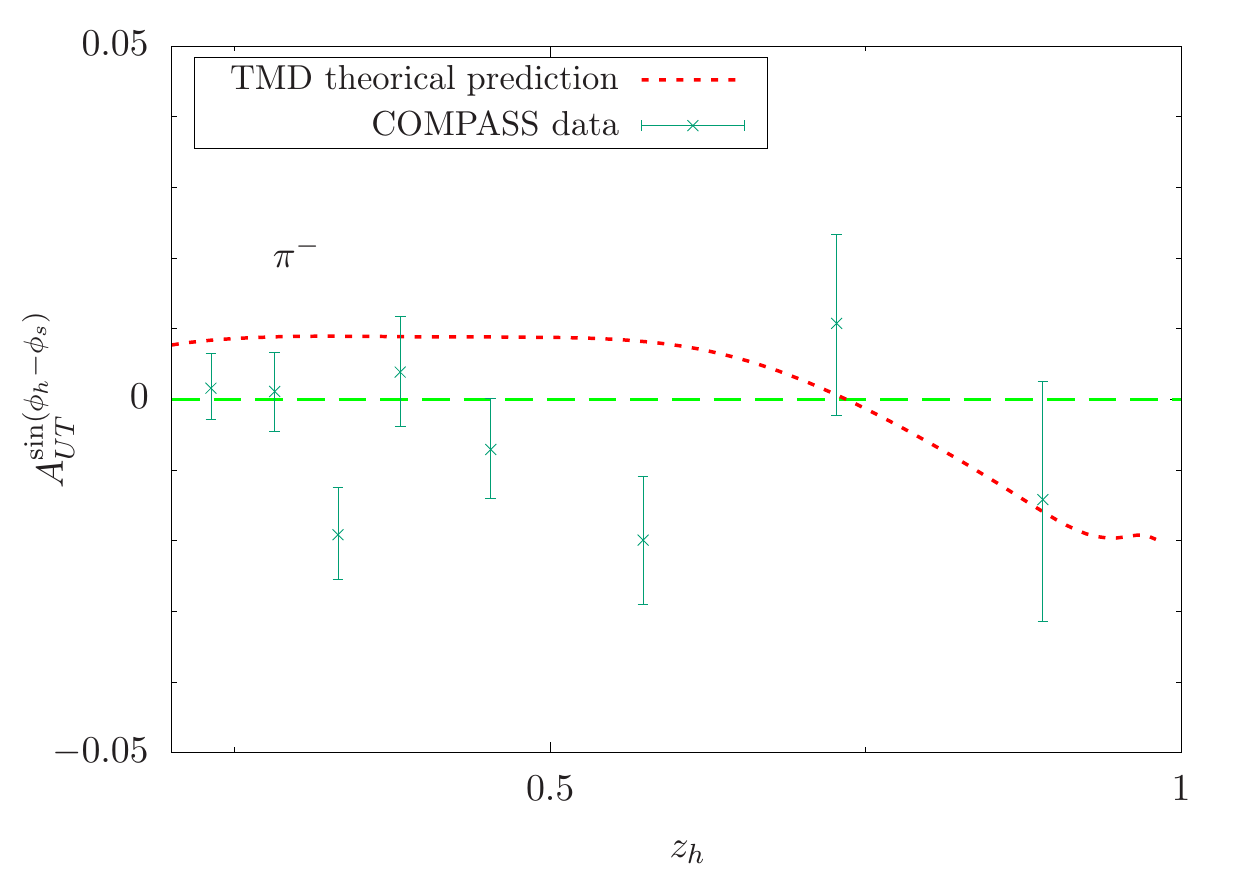}
\includegraphics[scale=0.46]{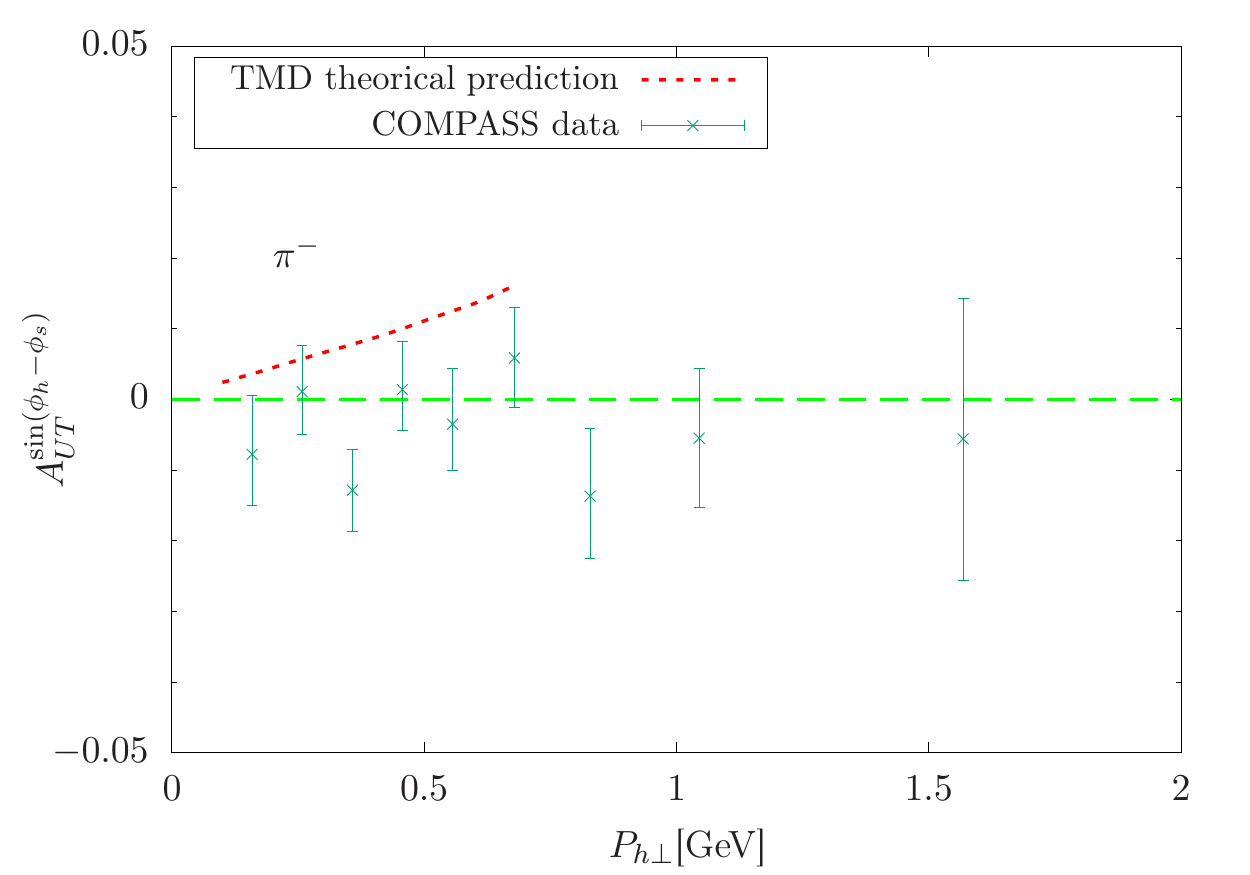}
\caption{ The Sivers asymmetry calculated within TMD factorization, compared with the COMPASS measurement \cite{Adolph:2014zba} for $\pi^-$ production.}
\label{fig8}
\end{figure}
\begin{figure}[htp]
\centering
\includegraphics[scale=0.46]{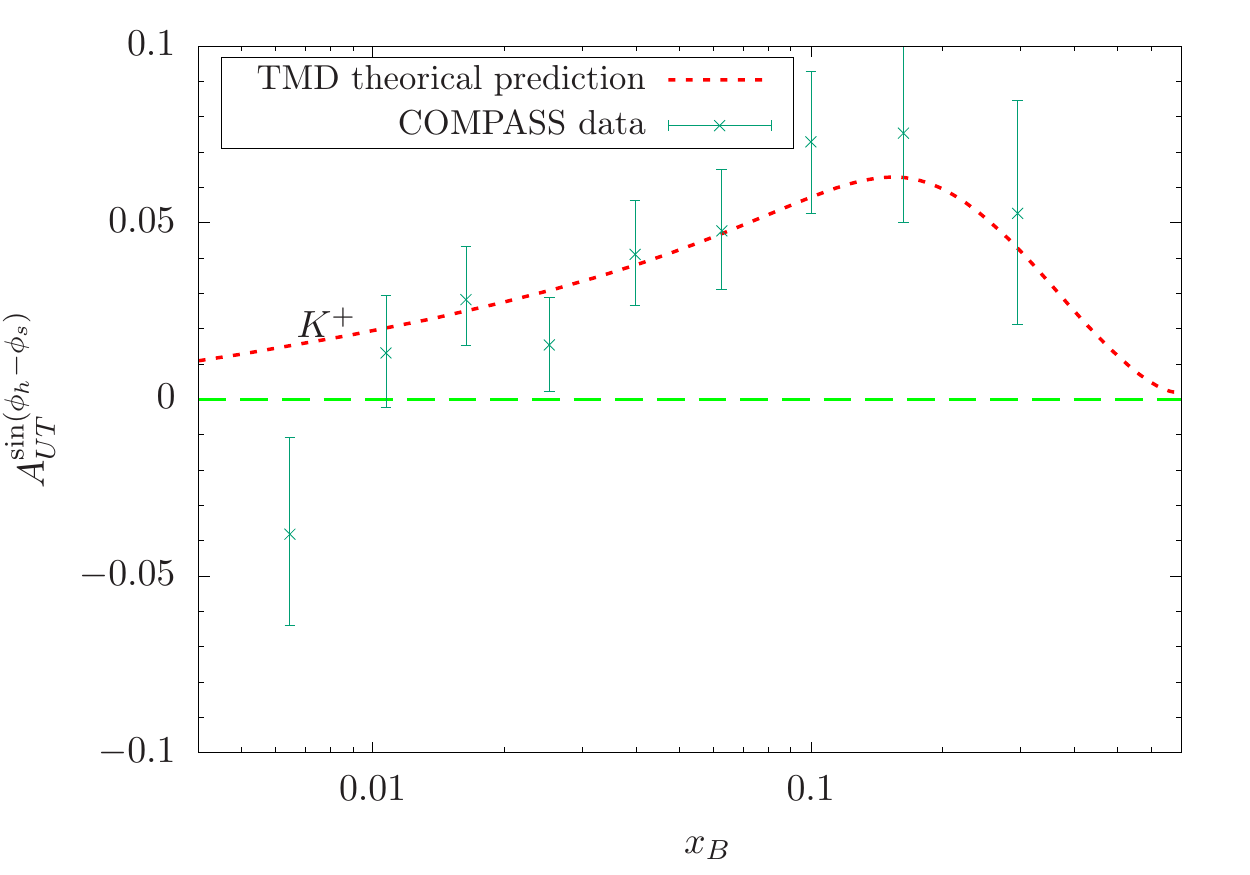}
\includegraphics[scale=0.46]{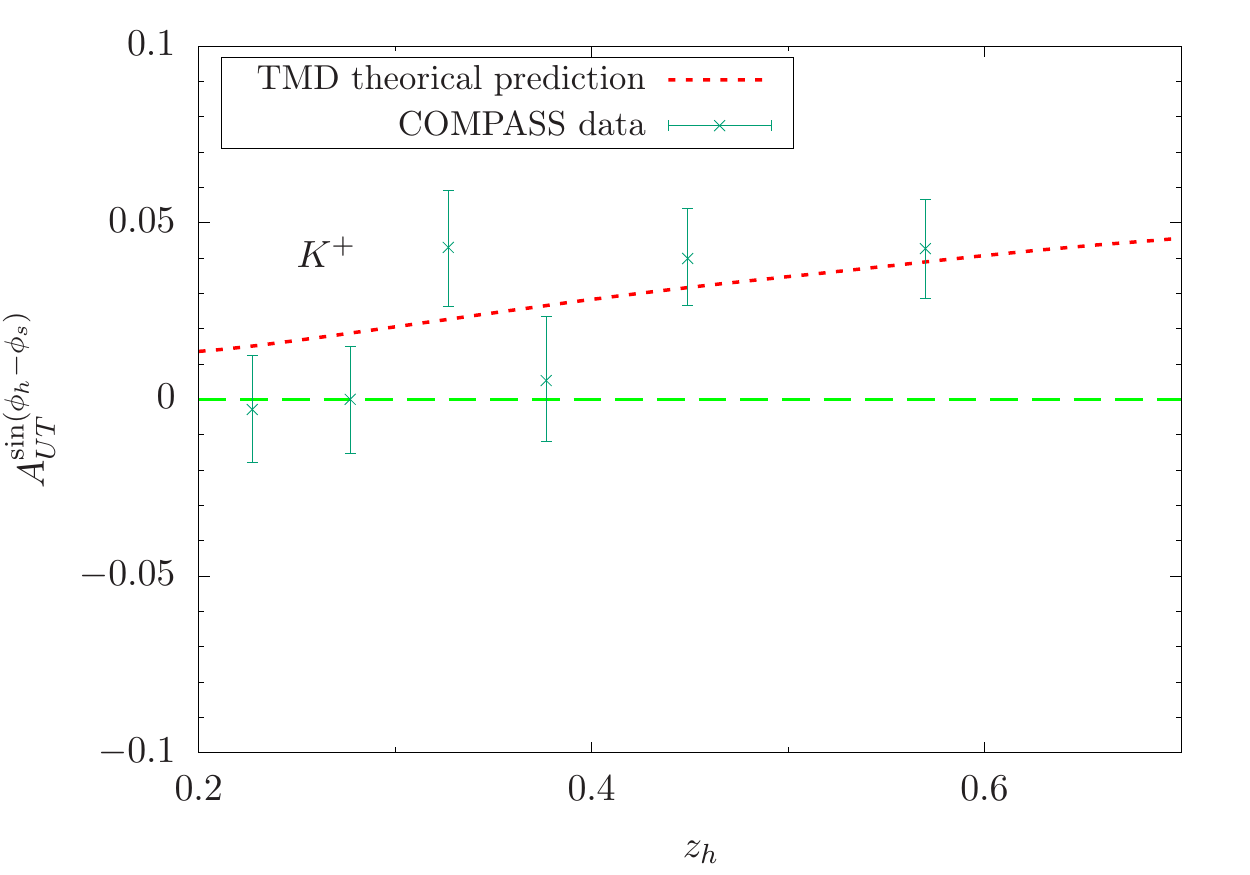}
\includegraphics[scale=0.46]{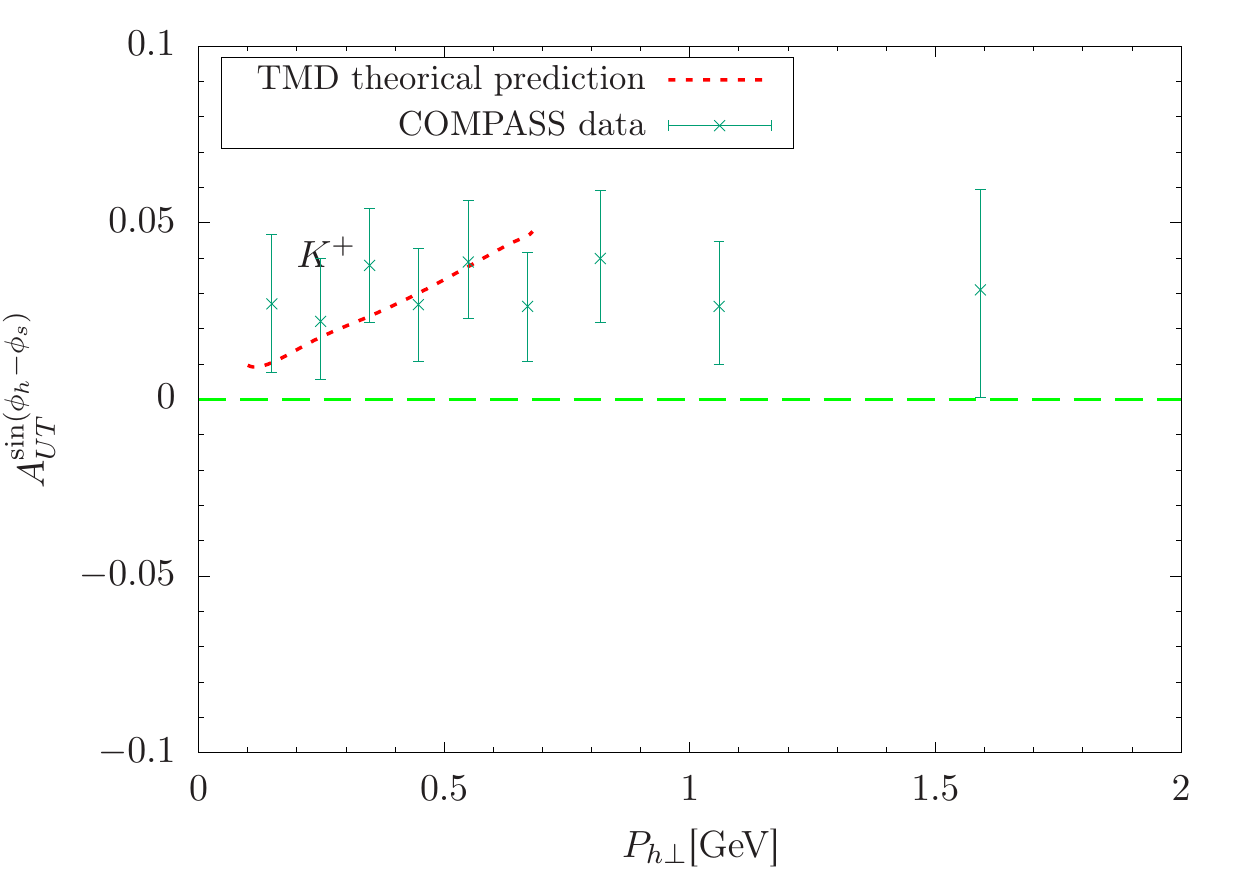}
\caption{ The Sivers asymmetry calculated within TMD factorization, compared with the COMPASS measurement \cite{Adolph:2014zba} for $K^+$ production.}
\label{fig9}
\end{figure}
\begin{figure}[htp]
\centering
\includegraphics[scale=0.46]{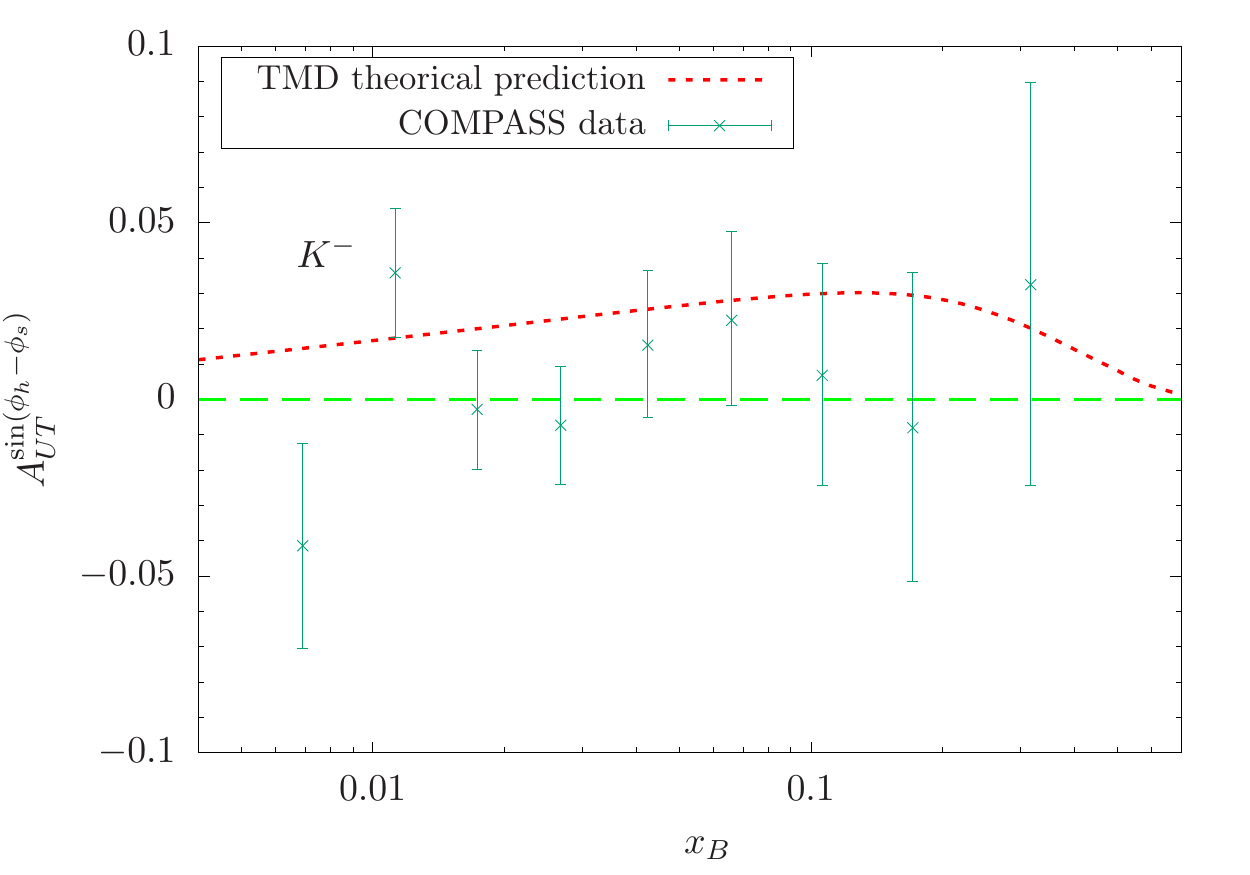}
\includegraphics[scale=0.46]{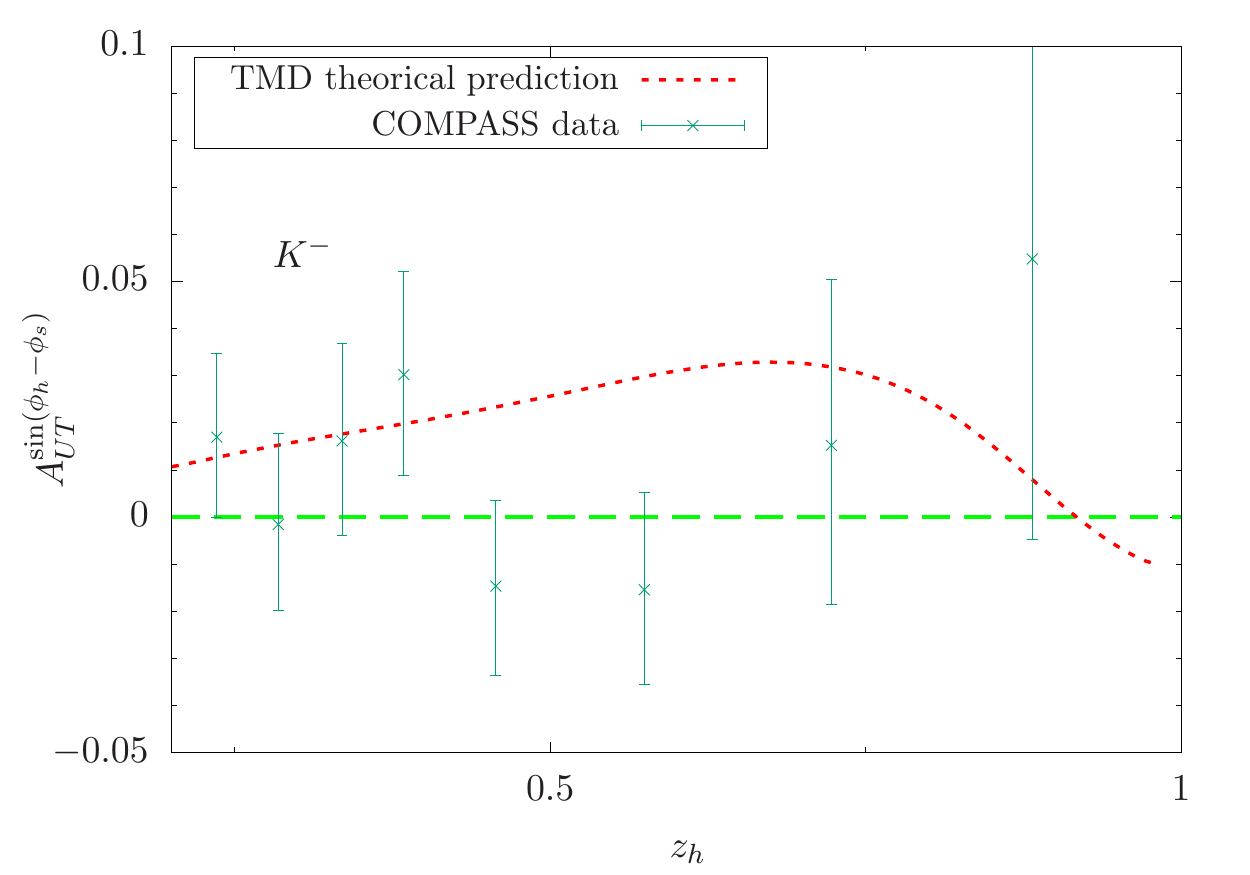}
\includegraphics[scale=0.46]{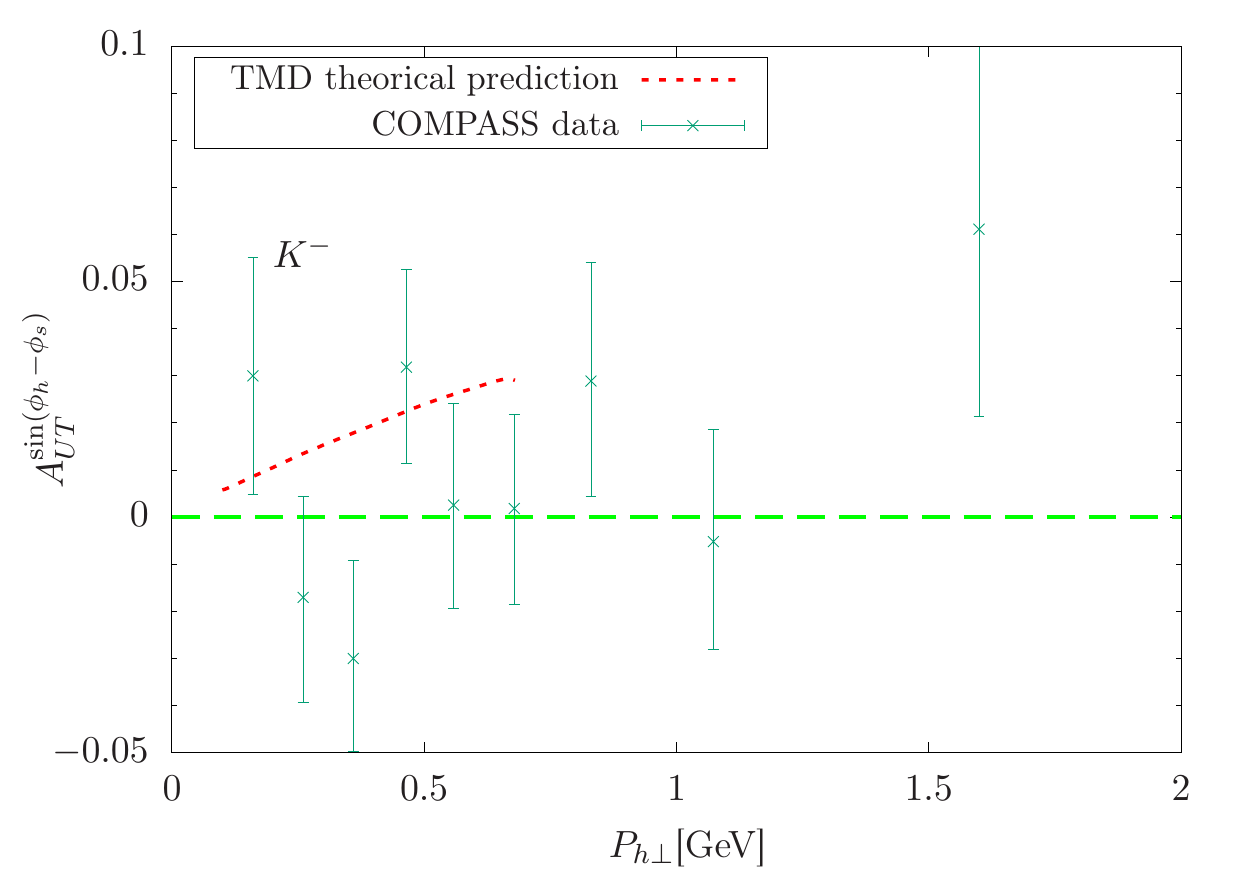}
\caption{ The Sivers asymmetry calculated within TMD factorization, compared with the COMPASS measurement \cite{Adolph:2014zba} for $K^-$ production.}
\label{fig10}
\end{figure}
We also make predictions for Sivers asymmetries at COMPASS, with a muon beam of 160 GeV scattered off a proton target.
The kinematical cuts we employ in the calculation are \cite{Adolph:2014zba}
\begin{eqnarray} \label{eq48}
\begin{aligned}
&0.004 <x_B< 0.7 \qquad 0.1<y<0.9 \qquad 0.2<z_h<1 \qquad P_{h\perp} > 0.1\text{GeV}
\\
&Q^2>1\text{GeV}^2 \qquad W>5\text{GeV}
\end{aligned}
\end{eqnarray}
The obtained $x_B$-, $z_h$-, and $P_{h\perp}$-dependent asymmetries for pion and kaon production are compared with the COMPASS data in Figs.\ref{fig7}-\ref{fig10}.
As shown in Figs.\ref{fig7} and \ref{fig9} ,in all cases, the asymmetries for $\pi^+$ and $K^+$ production acquired from our calculations are positive, which is consistent with the COMPASS data, whereas, the $z_h$-dependent asymmetry for $\pi^-$ production in Figs.\ref{fig8} is positive in the region $z_h < 0.72$ and is negative in the region $z_h > 0.72$.
In conclusion, the Sivers asymmetries reached within the TMD factorization and evolution at the corresponding kinematics are basically consistent with the HERMES and COMPASS measurements.

\section{Conclusion}\label{secIV}

In this paper, we study the single spin asymmetry $A_{UT}^{\sin(\phi_h-\phi_S)}$ of a single hadron production in SIDIS within the framework of TMD factorization up to NLL order of QCD.
We work out the energy evolutions of the Qiu-Sterman function by taking the parametrization at an initial energy $Q_0$ and evolving it to another energy $\mu_b$ through an approximation evolution kernel for the Qiu-Sterman function, including only the homogenous terms. Similarly, the timelike evolution of the unpolarized fragmentation function is also performed by QCDNUM.
Then we reach the $x_B$-, $z_h$-, and $P_{h\perp}$-dependent Sivers asymmetries for the pion and kaon production at the kinematics of HERMES and COMPASS experiments, respectively. The results are compared with the corresponding HERMES and COMPASS measurements. It is found that most of the Sivers asymmetries reached are basically consistent with the HERMES and COMPASS measurements.
However, there are still some reached Sivers asymmetries (e.g., in the three panel of Fig.\ref{fig8} )  that compare not so well with experimental data.

\begin{acknowledgments}
Hao Sun is supported by the National Natural Science Foundation of China (Grant No. 11675033) and by the Fundamental Research Funds for the Central Universities (Grant No. DUT18LK27).
\end{acknowledgments}

\bibliography{v1}

\end{document}